\documentclass{aa} 
\usepackage{txfonts}
\usepackage{graphicx} %Include figure files
\usepackage{dcolumn}% Align table columns on decimal point
\usepackage{bm}% bold math
\usepackage{multirow}
\usepackage{rotating}
\begin{document}

   \title{Relative drifts and temperature anisotropies of protons and $\alpha$ particles in the expanding solar wind -- 2.5D hybrid simulations}

%   \subtitle{Limitations on the differential streaming of $\alpha$ particles in the expanding solar wind}

   \author{Y. G. Maneva
          \inst{1,2}
          \and
          L. Ofman\inst{2,3}
          \and
         A. Vi{\~n}as\inst{3}
          }

   \institute{Centre for mathematical Plasma Astrophysics, KU Leuven, Celestijnenlaan 200B, B-3001 Leuven, Belgium\\
              \email{yana.maneva@wis.kuleuven.be}
         \and
            Department of Physics, Catholic University of America, 620 Michigan Ave NE, Washington D.C. 20064, USA \\
             \email{leon.ofman@nasa.gov}
         \and
         NASA Goddard Space Flight Center, 8800 Greenbelt Rd, Greenbelt, Maryland 20771, USA\\
            \email{adolfo.vinas@nasa.gov}
             }

   \date{Received: June 13, 2014\\ 
         Accepted: }

\titlerunning{Relative drifts and temperature anisotropies of $\alpha$ particles in SW}

  \abstract  
  % context heading (optional)
  % aims heading (mandatory)
  % methods heading (mandatory)
  % results heading (mandatory)
{We perform 2.5D hybrid simulations to investigate the origin and evolution of relative drift speeds between protons and $\alpha$ particles in the collisionless turbulent low-$\beta$ solar wind plasma.}
{We study the generation of differential streaming by wave-particle interactions and absorption of turbulent wave spectra. Next we focus on the role of the relative drifts for the turbulent heating and acceleration of ions in the collisionless fast solar wind streams.}
{The energy source is given by an initial broad-band spectrum of parallel propagating Alfv\'en-cyclotron waves, which co-exists with the plasma and is self-consistently coupled to the perpendicular ion bulk velocities. We include the effect of a gradual solar wind expansion, which cools and decelerates the minor ions. This paper for the first time considers the combined effect of self-consistently initialized dispersive turbulent Alfv\'enic spectra with differentially streaming protons and $\alpha$ particles in the expanding solar wind outflows within a 2.5D hybrid simulation study.}
{In the non-expanding wind, we find a threshold value of the differential streaming $V_{\alpha p} = 0.5 V_\mathrm{A}$, for which the relative drift speed remains nearly steady. For ions, streaming below the threshold value, the waves act to increase the magnitude of the relative drift speed. Ions, which stream faster than the threshold value become subject to nonlinear streaming instability and as the system evolves their bulk velocities decrease. We find that the solar wind expansion strongly affects the relative drift speeds and significantly slows down both ion species for all values of the relative drift speeds considered in this study.}
{}

   \maketitle

\section{Introduction}
Remote sensing and in situ measurements in coronal holes and fast solar wind show that the heavy ions are preferentially heated and accelerated, despite the higher inertia and the stronger gravitational force, which they experience in the vicinity of the solar surface \citep{Marsch:82a, Kohl:98,CFK99, Kohl:06}. In situ observations of the temperature anisotropies and the relative drift speed between protons and $\alpha$ particles in collisionless fast solar wind streams indicate a relation between the wave activity and the bulk properties of the ion species in the range $0.3-1 AU$ \citep{Bourouaine:10,Bourouaine:11,Maruca:11,Maruca:12,Kasper:08,Kasper:13,Bourouaine:13a}. Recent observations indicate that doubly-ionized solar wind helium ions near the Earth can be nearly 7 times hotter than protons on average \citep{Kasper:13}. What causes such a strong heating for the $\alpha$ particles and what leads to the onset of preferential acceleration and the generation of the observed temperature anisotropies are still open questions, remaining to be solved by future observations and models. Currently there are no measurements of helium ions in the inner solar corona, but spectroscopic observations of other ions like  $O^{5+}$ and $Mg^{9+}$ in coronal holes imply high outflow speeds, temperatures and anisotropies for the heavy species there \citep{Kohl:97,Kohl:99,CFK99,Ofman:01,Ofman:11,Ofman:13}. This indicates that the onset of the preferential heating, differential acceleration and temperature anisotropies for the heavy ions takes place in the low-$\beta$ coronal plasma, where turbulence is important and wave amplitudes can be higher than at 0.3 AU \citep{Jian:09,Jian:10}.
%by the solar physics community.

In this paper we perform 2.5D hybrid simulations to investigate the generation of ion temperature anisotropies and differential streaming by broadband turbulent spectra of low-frequency Alfv\'en-cyclotron waves. Further on, we study the influence of ion relative drift speeds and the solar wind expansion on the ion heating and differential acceleration. The results of this study provide insight into the physical mechanisms responsible for the observed preferential heating and acceleration of minor ions in the solar wind. The work relates the particles properties to the existing solar wind wave spectra, in terms of generation and absorption of waves via nonlinear wave-wave and wave-particle interactions, and the evolution of the micro-turbulence in the solar wind.

The scope of our paper is of broad interest to the solar plasma community and extends various previous studies based on hybrid numerical modeling, e.g., \citet{Gary:03,Gary:06,Xie:04,Ofman:05,OV07,Hellinger:06b,Perrone:13,Hellinger:13a,OfmanVM:14}, linear Vlasov kinetic instability theory, \citet{Gary:01b,Gary:02,Verscharen:13b,Chandran:13} and numerous observations, such as \citet{Reisenfeld:01,Gary:02,Hellinger:06a,Kasper:08}; \citet{Bourouaine:10,Bourouaine:11,Maruca:12,Kasper:13,Bourouaine:13a}. \citet{Podesta:11b} perform linear theory analysis to explore the effects of the $\alpha$-proton differential streaming in the solar wind. \citet{Hellinger:06b,Hellinger:13a} discuss the role of $\alpha$-proton relative drifts for destabilizing the plasma in nonlinear 2.5D hybrid simulations. \citet{Verscharen:13b}, \citet{Chandran:13} and \citet{Bourouaine:13a} consider the limits of differential streaming and ion temperature anisotropies from the point of view of linear kinetic instabilities and observations. 
\citet{Hellinger:13a} have studied the effects of the solar wind expansion on the plasma instabilities in high-plasma-$\beta$ system, consisting of differentially streaming proton core, beam and $\alpha$ particles in the heliosphere at distance $r > 0.3 AU$. In the current paper we investigate the effect of the solar wind expansion for an initially drifting $\alpha$-proton plasma in low plasma $\beta$ conditions appropriate in particular to the inner heliosphere and the solar corona. We initialize the system with an $\alpha$-proton population and let the ion beams be self-consistently generated by the model ambient wave-spectra as the system evolves in time. We load the particle populations with equal temperatures for the two ion species and study the temporal evolution of their temperature ratio as a function of the relative drift speed. With the goal of constructing a more realistic model of the solar wind, we consider the effects of turbulent initial wave spectra, which obey the three-species electron-proton-$\alpha$ cold-plasma dispersion relation. This allows for additional wave-particle and wave-wave interactions, non-present in the absence of waves, and provides essential information on the relation between plasma waves and the properties of ions in collisionless low-$\beta$ solar wind.
Recently, \citet{OfmanVM:14} have considered the effect of the solar wind expansion for the cases of super-Alfv\'enic initial relative drifts or $\alpha$ particle temperature anisotropies, and a driven wave spectra. The waves are injected in time at one side of the periodic boundaries with a prescribed power law consistent with observations within a certain range in frequency space. While the study of \citet{OfmanVM:14} is based on the introduction of time-dependent magnetic fluctuations, in the present study the magnetic spectrum is introduced as an initial value problem. By construction, the broad-band spectra preserves the magnetic field divergence-free, conserves the sense of magnetic helicity and obeys Parseval's theorem for energy conservation. The waves represent solutions of the cold plasma dispersion relation for a given narrow range in wave-vector space, for which a $k^{-1}$ spectral slope is considered.
Previous 1.5D and 2.5D hybrid simulations \citep{Xie:04,Araneda:09,OV07,OVM:11,Maneva:13a,Maneva:14,OfmanVM:14} show that monochromatic pumps and broad-band spectra of waves can preferentially heat and accelerate the $\alpha$ particles. In this 2.5D study we have narrowed the spectral range of the initial turbulent fluctuations in order to avoid the slow evolution of the very low-frequency MHD-type Alfv\'en waves, which would not interact with the particles within the simulation time-scales considered here (hundreds of gyroperiods).
Details of the simulation setup and the initial wave-spectra are presented in Section~2. 
Simulation results are described in Section~3 and conclusions are given in Section~4.

\section{Simulation setup}

In this section we introduce the normalization units and the characteristic parameters used for the simulations. The expanding box model and the construction of the initial wave spectra are presented in separate subsections.\\
The simulation time is given in units of proton gyro-frequency $\Omega_p^{-1}$ and the velocities are normalized to the local Alfv\'en speed, $V_\mathrm{A} \equiv \frac{B_0}{\sqrt{\mu_0 n_em_p}},$ as defined by the magnitude of the homogeneous background magnetic field $B_0$, the electron number density $n_e$ and the proton mass $m_p$. The length of the simulation box is the same in both spatial directions. In units of the proton inertial length it reads $L_x = L_y = 384 V_{\mathrm A}/\Omega_p.$ The simulations are performed with 256$\times$256 cells and 127 particles per cell per species, or in total $> 16.6$ million particles. The time step used is a small fraction of the proton gyroperiod, $\Delta t = 0.02 \Omega^{-1}_p$. 

\subsection{Initial wave spectra}

\begin{figure}
\centering
\noindent\includegraphics[width=0.48\textwidth]{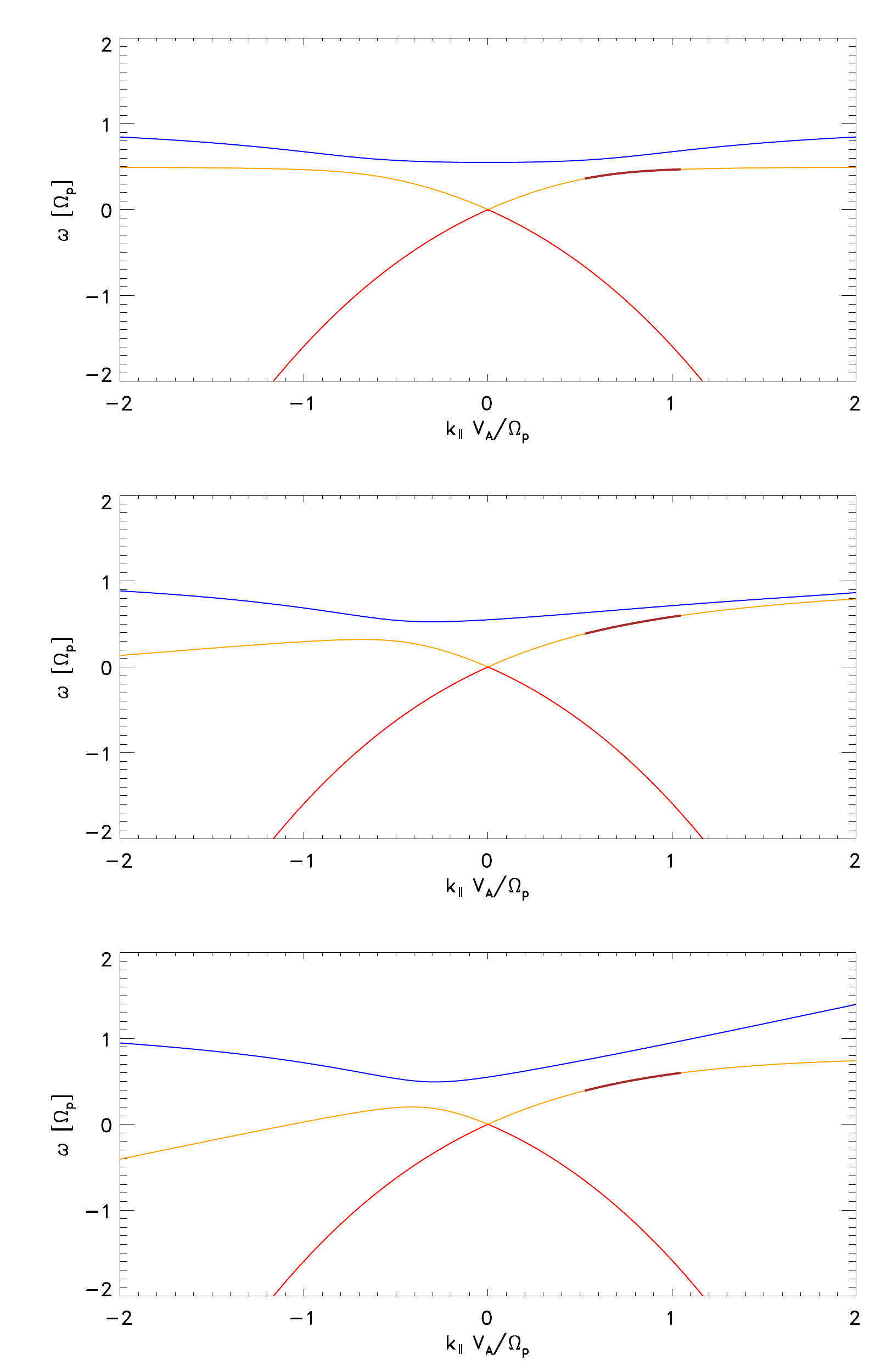}
\caption{Initial parallel propagating Alfv\'en-$\alpha$-cyclotron waves spectra -- solution of the coupled set of cold quasi-neutral multi-fluid plasma  and the non-relativistic Maxwell equations. The top panel shows the solution for non-drifting plasma. The middle panel shows the solution for relative drift speed $V_{\alpha p}=0.2 V_{\mathrm A}$. The bottom panel shows the solution for the special case of relative drift speed $V_{\alpha p}=0.5 V_{\mathrm A}$. The initial wave spectra was selected from the Alfv\'en-$\alpha$-cyclotron branch and its $[\omega_0,k_0]$ range is shown in solid brown.}
\label{fig:sonn_dispe}
\end{figure}

To initialize the simulations we assume that the magnetic field consists of a constant background and a fluctuating part, which corresponds to a broad-band spectrum of low-frequency parallel propagating Alfv\'en-cyclotron waves:
\begin{equation}
\mathbf{B} = \mathbf{B_0} + \mathbf{B_\perp}.
\end{equation}
To construct the initial spectra for the two-dimensional simulations we perform a one-dimensional parallel propagating broad-band reconstruction, based on the procedure described in \citet{VinasMAM:14}, and distribute it homogeneously at each point in the transverse direction until the entire simulation domain is filled with waves. 
For the one-dimensional reconstruction we assume that the ambient magnetic field is along the $x$ axis, $\mathbf{B_0} = B_0 \hat{x}$, and the fluctuating part $\mathbf{B_\perp}(x)$ represents left-hand polarized transverse waves propagating along the background field. For the magnetic fluctuations we consider plane waves decomposition:
\begin{equation}
%\mathbf{B_\perp}= \textit{Re}\[\sum_{k>0} (C_k^L \exp(i\phi_k) (\hat y + i \hat z) + C^R_k \exp(i\psi_k) (\hat y − i \hat z))\exp(ikx) \],
\mathbf{B_\perp}= \textit{Re}\left[\sum_{k>0} C_k^L \exp(i\phi_k) (\hat{y} + i \hat{z})\exp(ikx) \right],
\end{equation}
where the phases $\phi_k$  are randomly set between $\pm [0, 2\pi]$. The Fourier coefficients for left-hand polarized waves $C_k^L$ are computed based on the following physical conditions: 1) a divergence-free magnetic field $\nabla \cdot \mathbf B=0$, 2) preservation of the sense of magnetic helicity, and 3) conservation of energy between configuration and Fourier spaces, also known as Parseval's theorem. 
For left-hand circularly polarized waves the conservation of energy can be expressed as:
\begin{equation}
E = \frac{1}{2} \int {|B_\perp(x)|}^2 dx = \frac{1}{2} \textit{Re} \sum_{k>0} {|C^L_k|}^2.
\end{equation}
The total magnetic helicity for all fluctuations, defined as the scalar product of the vector potential $\mathbf A$ with the fluctuating magnetic field $\mathbf B$, can be represented as a superposition of the spectral magnetic helicity $H_{mk}$ for each mode
\begin{equation}
H_m = \frac{1}{2} <\mathbf A \cdot \mathbf B> = \frac{1}{2} \sum_k H_{mk}.% =\sum_kk \textit{Im}(A∗ykyk Azkzk)
\end{equation}
For a spectrum, which consists only of left-hand polarized waves, the spectral magnetic helicity can easily be expressed in terms of the Fourier amplitudes of the fluctuations as follows:
\begin{equation}
H_{mk} = \frac{1}{4k}\sum_k({|C^L_k|}^2).
\end{equation}
To construct a magnetic field power spectra, which resembles solar wind observations we assume a power-law dependence for the amplitude of the waves ${|C^L_k|}^2 = C^2 k^{-\alpha}$, where $\alpha$ is a prescribed power spectral index. The constant $C$, depends on the total amplitude of the magnetic field fluctuations $|\delta \mathbf B_\perp| = \sum_{k>0} {|C^L_k|}^2$, the spectral slope and the selected spectral range
\begin{equation}
C^2 = \frac{{\delta B}^2}{\sum_{k>0} k^{-\alpha}}.
\end{equation}
The spectra is constructed in Fourier space and inverse Fourier transform is used to provide the magnetic and velocity field fluctuations in configuration space. A detailed description of the method used to construct the initial broad-band spectra for a general combination of left-hand and right-hand circularly-polarized parallel propagating waves can be found in \citet{VinasMAM:14}. For the power spectral slope in our simulation study we choose $\alpha =-1$. We should note that the observed solar wind power spectra close to the Sun are in frequency domain. Thus the observed power spectral slope at large scales $\omega^{-1}$ can be directly related to $k^{-1}$ spectrum only for the low-frequency linear part of the dispersion relation, where $\omega=V_\mathrm{A}k$.\\
In order to self-consistently couple the initial ion velocities to the reconstructed magnetic field one needs to prescribe a certain dispersion relation for the waves and solve it to find the relation between the ion motion and the prescribed electromagnetic fluctuations. Complementary to the wave-spectra initialization in our previous 1.5D work \citep{Maneva:13a}, in the present 2.5D study we construct the initial ion bulk velocity fluctuations as solutions of the 3-fluid cold plasma dispersion relation, see Eq.~\ref{pumpdispe} and Fig.~\ref{fig:sonn_dispe}. In other words, we take into account the fact that the given spectrum of parallel propagating Alfv\'en-cyclotron waves imposes separate transverse bulk velocity fluctuations for the protons and the $\alpha$ particles, cf. with Eq.~\ref{initdrift}. Using a consistent initialization for both ion species is important to avoid artificial pressure imbalances and instabilities, which can affect the resulting ion heating and acceleration. We should note that for the chosen range of wave-numbers in the low plasma $\beta$ regime considered here (see Table~1), the solution of the cold plasma dispersion relation for parallel wave propagation in drifting isotropic plasma is practically identical to the solution of the full Vlasov linear theory dispersion relation, as there is no damping for the Alfv\'en-cyclotron branch at wave-numbers $k_0 < 0.5\Omega_p/V_{\mathrm A}$. Thus the chosen spectra are stable with respect to linear plasma instabilities, which become important at higher wave-numbers $k_0 > 0.5 \Omega_p/V_{\mathrm A}$, as shown for example by \citep{OV07}. 
To derive the cold plasma dispersion for parallel wave propagation, we start with the combined set of multi-fluid-Maxwell equations, which govern the behavior of low-frequency waves and particles in a drifting isotropic plasma:
\begin{eqnarray}
\frac{\partial n_s}{\partial t} + \nabla \cdot (n_s \mathbf V_s) = 0 \\
\frac{\partial \mathbf V_s}{\partial t} + \mathbf V_s \cdot \nabla \mathbf V_s - 
\frac{q_s}{m_s}(\mathbf E + \mathbf V_s \times \mathbf B) + \frac{\nabla P_s}{m_sn_s} =0 \\
\nabla \times \mathbf E = - \frac{\partial \mathbf B}{\partial t} \\
\nabla \times \mathbf B = - \mu_0 \mathbf J. 
\end{eqnarray}
The solution of this system for parallel wave-propagation is given by the dispersion relation \citep{Smith:64,Sonnerup:67, Araneda:09}:
\begin{equation}
\label{pumpdispe}
k_0^2 - \frac{\mu_0}{B^2_0} \sum_{i\neq e}n_im_i\frac{{\left(\omega_0 - V_ik_0\right)}^2}{1-\omega_0/\Omega_i + V_ik_0/\Omega_i}=0,
\end{equation}
where quasi-neutrality and current conservation are assumed to eliminate the electron contribution and the summation is carried over the ions. Since we are interested in low-frequency waves in non-relativistic plasma, we have neglected the vacuum electromagnetic fluctuations described by the $\omega_0^2/c^2$ term. The dispersion relation for warm drifting multi-species plasma can be found for example in \citep{DO75,Gar93,XOV04}.
For the velocity fluctuations of the ion species we obtain the relation:
\begin{equation}
\label{initdrift}
\mathbf V_{\perp i} = -\frac{\omega_0/k_0 - V_i}{1-(\omega_0 - V_ik_0)/\Omega_i} \delta \mathbf B_\perp/B_0,
\end{equation}
where $V_i$ is the bulk velocity speed for each species, which determines the differential streaming between the protons and the $\alpha$ particles: $V_{\alpha p} = V_\alpha - V_p.$ The velocity and the magnetic field fluctuations are coupled, so that the initially isotropic ions acquire a bulk transverse motion, determined by the properties of the wave.
%, see Eq.~\ref{pumpdispe} and Fig.~\ref{fig:sonn_dispe},
When no drifts are considered the corresponding velocity fluctuations for the $\alpha$ particles are much higher than the ones for the protons for the same given choice of initial parallel Alfv\'en-cyclotron waves, see e.g., \citet{Ofman:05}. In the presence of relative drifts the bulk velocity fluctuations for the protons and the $\alpha$ particles change. For $V_{\alpha p} = 0.2 V_\mathrm A$ the transverse velocity fluctuations for the 2 species become of the same order and for $V_{\alpha p} = 0.5 V_\mathrm A$ the proton velocity fluctuations are dominant.\\
The initial spectrum consists of 16 modes laying on the left-hand circularly-polarized Alfv\'en-$\alpha$-cyclotron branch (with a frequency below the $\alpha$-cyclotron frequency). As noted above, the spectra is chosen from the low-frequency solution of the dispersion relation (see Figure~1), which is stable with respect to warm plasma linear Vlasov instability theory and contains no significant wave amplification, nor damping. The selected normalized wave-number space given in terms of the ion inertial length as given by the local proton gyrofrequency $\Omega_p$ and the local Alfv\'en speed $V_{\mathrm A}$ is $k_0 \in [0.26-0.52] \Omega_p/V_{\mathrm A}$ or equivalently by the ratio of the proton plasma frequency to the speed of light $\omega_{p}/c.$ The corresponding frequencies for the individual runs are presented in Table~1. Note that in a previous 1.5D study, \citep{Maneva:13a}, we find that spreading the wave power within a rather wide range of frequencies and wave-numbers, including very low-frequency MHD-type non-dispersive waves, leads to significant reduction in the ion heating and particularly acceleration rate. This is due to the slow direct cascade and the low left-over power at the relevant kinetic scales. To investigate the ion heating and acceleration, in the present study we have considered initial wave spectra within the intermediate frequency range given above (in between the low-frequency MHD type Alfv\'en waves and the resonant ion-cyclotron waves), where dispersive effects facilitate nonlinear cascade and energy transfer towards the relevant scales for wave-particle interactions.
For this test study, the simulations are performed with a total magnitude for the entire initial wave spectra chosen to be $20\%$ of the magnitude of the external background magnetic field $|\delta \mathbf B_\perp| = 0.2 B_0$. If the constituent waves had the same phase (which is randomly selected in our model), this would translates to amplitudes of the individual waves of about $1 \%$ of the ambient magnetic field. Recent in situ observations of monochromatic Alfv\'en-cyclotron waves with similar amplitudes at 1AU have been presented in \citep{Jian:09}. The authors of that paper argue that larger amplitudes should be expected at smaller heliocentric distances, closer to the Sun. Large-amplitude magnetic fluctuations, presumably of Alfv\'enic nature, have also been detected by Helios at various distances in the inner heliosphere \citep{Bourouaine:11}. \textit{Wind} observations of large-amplitude Alfv\'en waves at 1AU have recently been reported in \citet{Wang:12}.
%MESSENGER at 0.28 AU (Gershman et al 2012). 
\subsection{Expanding box model}
The expanding box model used in our simulation setup follows the original work by \citep{Grappin:96} and \citep{Liewer:01}, which has been further utilized by many authors, for example \citet{Hellinger:05,Hellinger:06b,OVM:11,Hellinger:13a,Maneva:13a,OfmanVM:14}. The expanding box equations for a two-dimensional hybrid simulation setting read:
\begin{equation}
\frac{\mathrm{d} x^{'}}{\mathrm{d} t^{'}} = v_x^{'}, \quad \frac{\mathrm{d} y^{'}}{\mathrm{d} t^{'}} = \frac{1}{a(t)}v_y^{'}
%\quad \frac{\mathrm{d} z^{'}}{\mathrm{d} t^{'}} = \frac{1}{a(t)}v_z^{'},
\end{equation}
where $x$ is the direction of radial expansion and the external magnetic field, $a(t)=1 + U_0t/R_0$ and the ion velocities in the expanding (prime) frame and at rest are related by the magnitude of the constant solar wind bulk speed $U_0$ and the initial distance from the Sun $R_0$
\begin{equation}
\nonumber v_x^{'} = v_x - U_0, \quad v_y^{'} = v_y - U_0/R_0 y^{'}, \quad v_z^{'} = v_z- U_0/R_0 z^{'}.
\end{equation}
The electric field in the expanding (prime) frame of reference is computed from the momentum equation for the massless fluid isothermal electrons
\begin{equation}
\nonumber n_em_e \frac{\mathrm d \mathbf V^{'}_e}{\mathrm d t}= -en_e \mathbf E^{'} + \frac{(\nabla^{'}\times\mathbf B^{'})\times\mathbf B^{'}}{4\pi} - \frac{\mathbf{J^{'}_i}\times\mathbf{B^{'}}}{c} -\nabla^{'} P_e = 0,
\end{equation}
where $\mathbf{J_i^{'}}$ is the ion current in the co-moving reference frame, $P_e = n_ek_\mathrm{B}T_e$ is the electron pressure and the transformations for the electric field and the bulk velocities for all species are respecively $\mathbf E^{'}=\mathbf E + \mathbf{U_0} \times \mathbf{B}$ and $\mathbf{V_s^{'}}=\mathbf{V_s} - \mathbf{U_0}.$ 
The evolution of the magnetic field is modified by the expansion as follows
\begin{equation}
\frac{\partial \mathbf B^{'}}{\partial t}= - c (\nabla^{'} \times \mathbf E^{'}) - \frac{U_0}{R(t)} \hat{\mathbf L}\cdot \mathbf{B}^{'}, 
\end{equation}
where $\hat{\mathbf L} = (2,1,1)\delta_{ij}$ is the transformation matrix which accounts for the effect of the slow expansion, determined by the expansion parameter $\varepsilon t\equiv U_0/R_0 \ll 1,$ where $t$ is given in units of $\Omega_p^{-1}$. In the considerations above all second and higher order terms of the type $O(\varepsilon^2)$ have been neglected.
In the present hybrid model we solve the above system of equations conserving the net charge and the total current. The charge neutrality for the 3 species plasma is set by $n_e = n_p + 2n_\alpha$ and the simulations are initialized with zero net current, which is conserved throughout the dynamical evolution of the system. As an initial state we consider isotropic plasma with equal ion temperature for the protons and the $\alpha$ particles, $T_\alpha=T_p$, and different value of the parallel relative drift speed between the 2 species $V_{\alpha p}$. The parallel components of the corresponding ion plasma beta $\beta_{\parallel i} = v^2_{\mathrm{th},\parallel i}/ V^2_\mathrm{A},$ defined in terms of the parallel ion thermal speed $v_{\mathrm{th},\parallel i}$ and the local Alfv\'en speed $V_\mathrm{A}$ is $\beta_\alpha=0.02$ for the He$^{++}$ ions and $\beta_p=0.08$ for the protons. The $\alpha$ particles constitute $5\%$ of the background electron density, $n_\alpha = 0.05 n_e$ and the plasma beta of the isothermal fluid electrons is $\beta_e = 2\mu_0n_e k_\mathrm B T_e/B_0^2 = 0.05$. Most simulations are computed until $t= 700 \Omega_p^{-1}$, whereas the simulations with higher expansion factor (see Table~1) are computed until $t= 600 \Omega_p^{-1}$ to comply with the approximation used in the expanding box model, $\varepsilon t <<1$.

\section{Results}

Figure~\ref{fig:sonn_dispe} describes the shift of the dispersion relation branches due to the presence of relative drift speed between the protons and the $\alpha$ particles within the cold quasi-neutral plasma approximation considered here. The red line represents the right-hand polarized fast mode, excited by the electrons. The blue line corresponds to the left-hand polarized proton-cyclotron branch and the orange curve represents the left-hand polarized Alfv\'en-cyclotron waves, related to the heaviest $\alpha$ particles in the three-species plasma considered here. The selected initial wave-spectra belongs to the latter low-frequency Alfv\'en-alpha-cyclotron branch and is overplotted in brown color. The relative drift leads to asymmetries in the forward and backward propagating Alfv\'en-cyclotron branches and increases the forbidden bands between the proton- and the $\alpha$-cyclotron branches found in the linear dispersion relation. Note, that the relative drift has no visible influence over the initially selected frequency range for our simulations. Nevertheless, it significantly changes the bulk velocity fluctuations for the different ions, see Eq.~\ref{initdrift} and Figure~2. The relative drifts also influence the turbulent evolution of the initial wave spectra as shown later on Figures~10-13.\\
Figure~2 shows the Alfv\'enic correlation between the reconstructed initial ion bulk velocities and the magnetic field fluctuations as a function of the simulation box length in ``$x$'' direction, parallel to the background magnetic field. For the sake of simplicity the figure shows only one component of the transverse velocities (given in red) and magnetic field fluctuations (described by the blue lines). The other transverse components exhibit identical correlation. The left panel shows the correlation for protons and the right panel describes the $\alpha$ particles. The plots illustrate the change in the magnitude of the velocity fluctuations for the different ion species depending on the value of the initial relative drift speed. The top panels correspond to non-drifting distributions. The middle panels show the case with $V_{\alpha p}=0.2 V_\mathrm{A}$ and the lower panels stand for the case with $V_{\alpha p}=0.5 V_\mathrm{A}.$ The figure demonstrates how the transverse velocity fluctuations for the $\alpha$ particles decrease with increasing relative drift speed. Since the current conservation condition in a frame co-moving with the electrons implies $V_\alpha >> V_p$, the proton velocity fluctuations are in the opposite direction to the $\alpha$'s and remain nearly unaffected by the values of the relative drift considered here.\\
Figure~3 shows the temporal evolution of the ion temperature anisotropies and the relative drift speed. Overplotted with dashed lines is the effect of a gradual solar wind expansion with $\varepsilon t_0 = 10^{-4}$, where $t_0=1\Omega_p^{-1}$. The initial temperature anisotropy for both species is apparent and is caused by the non-thermal transverse velocity fluctuations, induced by the waves. The onset of the differential streaming between the protons and the $\alpha$ particles is caused by the acceleration of the proton beam in the system with zero net currents. The close to realistic expansion factor considered here leads to a minor change in the bulk ion properties.\\
Figure~4 and Figure~5 also describe the temporal evolution of the ion anisotropies and relative drift speed, but for the case of initially drifting distributions with $V_{\alpha p}=0.2 V_\mathrm A$ and $V_{\alpha p}=0.5 V_\mathrm A$, respectively. The effect of the gradual expansion varies with the value of the initial relative drift speed. It leads to less than $2\%$ change in the ion temperature anisotropies and a $4.5\%$ change in the relative drift speed for the case of $V_{\alpha p}=0.2 V_\mathrm A$. As we increase the initial relative drift speed, the solar wind expansion leads to about $17\%$ reduction for the temperature anisotropy of the $\alpha$ particles and a $8.9\%$ reduction of the proton temperature anisotropy, whereas the evolution of the relative drift speed stays almost unchanged.\\
In order to understand the relation between the initial wave spectra and the apparent temperature anisotropy of the ions we should take into account several factors. To begin with, the simulations are initialized with equal temperatures for both ion species. The normalized parallel ion temperature for each species is set by half the product of their mass ratio and the parallel plasma $\beta$, i.e., $T_{\parallel, i}=m_i\beta_{\parallel ,i}/2m_p,$ so that $T_{\parallel, p}=T_{\parallel, \alpha}=0.04$. Since we assume initially isotropic plasma, the initial perpendicular temperatures for both ion species are also equal. What appears on the plots instead are the apparent temperatures and anisotropies, which are calculated based on the thermal velocities of the ions and the non-thermal ion motions, imposed by the initial wave spectra, see Eq.~\ref{initdrift}. Since we consider only parallel waves their effect on the ions is only in perpendicular direction. Therefore the larger non-thermal component of the velocity fluctuations induce higher apparent temperature anisotropies. This picture changes for oblique wave propagation, where the waves affect also the parallel motion of the ions and induce parallel bulk velocity fluctuations as well [Maneva et al., in preparation]. 
For the initial wave spectra used in our simulations the magnitude of the velocity fluctuations of the $\alpha$ particles and the protons constitutes from $42\%$ to almost twice the value of their initial perpendicular thermal speeds ($v_{\mathrm{th}, \perp \alpha} = \sqrt{0.02} V_\mathrm{A}$ and $v_{\mathrm{th}, \perp p} = \sqrt{0.08} V_\mathrm{A}$). The ratio of the average wave-induced perpendicular bulk velocity fluctuations to the thermal velocities of the ions at the beginning of the simulations as a function of the relative drift speed between the two species is presented in Table~1. At zero drifts the motion of the $\alpha$ particles is strongly affected by the wave-spectra, see Figure~2, and the average of their perpendicular velocity flucutations for the chosen broad-band spectra is $<V_{\perp \alpha}> = 0.26 V_\mathrm{A}$ vs. $<V_{\perp p}> = 0.15 V_\mathrm{A}$ for the protons. As the relative drift speed increases to $V_{\alpha p} =0.2V_\mathrm{A}$ the selected wave-spectra affects the $\alpha$ particles and the protons in a similar manner and the imposed velocity flucutuations for the two species are of the same order: $<V_{\perp \alpha}> \approx <V_{\perp p}> \approx 0.16 V_\mathrm{A}.$ For higher drifts, $V_{\alpha p} =0.5V_\mathrm{A}$, the influence of the electromagnetic field of the wave spectra on the $\alpha$ particles is negligible $<V_{\perp \alpha}> = 0.06 V_\mathrm{A},$ whereas its effect on the protons remains almost unchanged $<V_{\perp p}> = 0.17 V_\mathrm{A}.$ We should note that for a fixed type of waves the relation between the velocity fluctuations and the relative drift speed strongly depends on the selected spectral range, as a shift in $\omega-k$ space would change the resonant condition according to Eq.~\ref{initdrift}. This relation is only valid for parallel wave propagation within the cold isotropic homogeneous plasma limit and it would change once oblique waves and anisotropic ions or plasma inhomogeneities are introduced.\\
Figures~6--8 describe the temporal evolution of the parallel and perpendicular temperature components for protons and $\alpha$ particles for the three sets of values for the initial relative drift speed, starting with a non-drifting distributions and increasing the drift to $V_{\alpha p}=0.2 V_\mathrm{A}$ and $V_{\alpha p}=0.5 V_\mathrm{A}.$ Each figure consists of two panels which demonstrate the difference between the ion heating and cooling in the non-expanding wind, panel (a), and the effect of a gradual expansion with $\varepsilon t_0 = 10^{-4}$, panel (b). The perpendicular component of the ion temperature is given by the solid red lines, whereas the parallel component is plotted with dash-dotted blue lines. The expected perpendicular cooling as predicted by the double-adiabatic CGL model is described by the solid green lines. In all cases the expansion leads to significant perpendicular cooling for both ion species and minor cooling in parallel direction. We observe about 16-20\% decrease in the perpendicular temperature for the $\alpha$ particles and about 6-11\% decrease in the perpendicular temperature for the protons. The parallel cooling is less prominent for both protons and $\alpha$ particles, and it decreases with increase of the initial relative drift speed. Thus the highest reduction of the ion temperature anisotropy for the $\alpha$ particles due to the effect of the gradual expansion is at $V_{\alpha p}=0.5 V_\mathrm{A}$, refer to Figure~5. As discussed above, the presence of relative drifts decreases the non-thermal component of the transverse velocity for the minor ions, which reduces their apparent perpendicular temperature. Despite the substantial perpendicular cooling caused by the expansion, the perpendicular temperature decrease for the minor ions is less than what is the expected by the CGL model and preferential heating for the $\alpha$ particles is present in all simulation cases. The protons on the other hand are less affected by initial and induced waves throughout the simulations, and their evolution is close to the double-adiabatic expectations.\\
In Figure~9 we show the contour plot of the two-dimensional ion velocity distribution function at the final stage of the simulations for the case of initially drifting plasma with $V_{\alpha p} =0.2 V_\mathrm{A}$. The parallel velocity component is along the $x$ axis and the perpendicular one is along the ordinate. The ion acceleration and anisotropic heating lead to deviation from the initial isotropic distributions. Ion acceleration associated with prominent ion beam formations occurs for both species. The beams are caused by the turbulence-generated ion-acoustic fluctuations, similar to the parametric instability scenario described in \citet{Araneda:09,Maneva:09}. The perpendicular heating is a combination of the quasilinear diffusion in phase-space from the forced bouncing caused by the initial wave-spectra and additional resonant scattering with the ion-cyclotron waves generated during the non-linear evolution of the system. To facilitate the comparison of their distributions, in this figure we use the same isotropic range in $v_x$ and $v_y$ for both protons and the $\alpha$ particles. The 2D velocity distributions for both ion species are symmetric in the phase-space formed by the transverse velocity components $v_y$ and $v_z,$ due to the circular polarization of the waves and the ion gyro motion. This makes the particle velocity distributions in $v_x,v_y$ and $v_x, v_z$ identical. The velocity distribution functions, corresponding to the non-drifting plasma case, show similar features for both ion species. However, in this case the proton beams are stronger and the $\alpha$ particles are heated to higher temperatures in both parallel and perpendicular direction. The stronger proton beams and parallel minor ion heating is likely related to Landau damping of the turbulence-generated ion density fluctuations, whose resonant absorption depends on the relative drift speed.\\
Figure~10 and Figure~11 show the power spectral density of the magnetic field fluctuations in Fourier space as a function of frequency and wave-numbers. Figure~10 corresponds to a non-drifting non-expanding plasma, whereas Figure~11 refers to a drifting non-expanding plasma with $V_{\alpha p}=0.2V_\mathrm A$. The top panels describe the redistributed wave power for the duration of the simulations as a function of the parallel wave-number, as projected along the ambient magnetic field. In both simulation cases, with and without relative drifts, the power spectra is asymmetric with respect to the direction of propagation. There are strong forward-propagating ion-cyclotron modes at $k_\parallel \leq 0.35 \Omega_p/V_\mathrm A$, remnants from the initial broad-band spectra, together with powerful forward-propagating fast modes, which are excited in the course of evolution. The initial waves with $k_\parallel > 0.35 \Omega_p/V_\mathrm A$ quickly decay and are absorbed by the plasma. A broad-spectra of powerful backward propagating higher order proton-cyclotron and some $\alpha$-cyclotron modes are also generated and intersect with the backward-propagating Alfv\'en-$\alpha$-cyclotron branch. The parallel dispersion relation is calculated by fixing a certain point in perpendicular $y$ direction and making a fast Fourier transform of the magnetic field fluctuations in time $t$ and along the parallel spatial coordinate $x$. The transverse dispersion relation is shown on the bottom panels of Figures~10-11. It is calculated following a similar procedure, but the fast Fourier transform was performed in $t$ and $y$, for a fixed point in $x$. The perpendicular power spectra indicate the presence of oblique modes, propagating across the ambient magnetic field. The oblique waves have higher intensity in the non-drifting case. The magnetic power spectra for both figures are computed over a large time interval, starting at the beginning of the simulations at $t=0$ and ending towards the end of the simulations at $t=600\Omega_p^{-1}$. In this way the plots illustrate the turbulent evolution of the fluctuations and the wave coupling at the fully nonlinear stage of the simulations. \\
Figure~12 and Figure~13 show the power spectral density of the magnetic field fluctuations (wave power) as a function of the parallel and perpendicular wave numbers at the end of the simulations for the same cases as described in Figures~10-11 above.
These complementary plots show the nonlinear anisotropic cascade of wave energy from parallel towards perpendicular wave-numbers.
The two-dimensional spectra are computed with a two-dimensional Fourier transform (along both spatial coordinates) of the fluctuating component of the magnetic field at the end stage of the simulations. The figures demonstrate the role of the differential streaming for the perpendicular energy cascade and oblique mode generation. The initial state for both cases is a broad-band spectrum of parallel Alfv\'en-cyclotron waves with parallel wave-numbers within the range $k_0 = k_\parallel = [0.26-0.52]\Omega_p/V_\mathrm{A}$. In the case of non-drifting plasma the initial spectrum is strongly depleted above $k_\parallel \leq 0.35 \Omega_p/V_\mathrm{A}$ and parallel waves at larger wavelengths are being generated, so that the initial spectra gets shifted towards lower parallel wave-numbers. In the drifting plasma case, the initial spectra for $k_\parallel > 0.35 \Omega_p/V_\mathrm{A}$ is not fully depleted and the generated oblique waves are slightly stronger. For both cases the strongest parallel waves are concentrated within $|k_\parallel| \in [0.1-0.3]\Omega_p/V_\mathrm{A}$, whereas the power of the generated oblique waves is concentrated within the limited range $|k_\perp| \in [0-0.1]\Omega_p/V_\mathrm{A}.$
Should we think of the spectra as a superposition of individual monochromatic waves, then we could try to understand the turbulent evolution in terms of superposition of the daughter waves generated by the parametric instabilities of the individual constituent pump waves. If one makes such an analogy, we could say that there are strong modulational and decay instabilities \citep{Kauffmann:08,Maneva:09}. There is an interplay between the modulational and the decay-like instabilities in the non-drifting case and the drifting case. We should note that the analogy with linear superposition of parametrically unstable waves is given only as an illustration of a possible way to introduce nonlinear effects. Realistic treatment of the solar wind turbulence requires fully nonlinear wave-wave couplings and consequent wave-particle interactions achieved in the present hybrid model.
\section{Discussion and Conclusions}

We have performed 2.5D hybrid simulations investigating the role of initially imposed broad-band wave spectra in a drifting and expanding solar wind plasma. We have studied the relaxation and evolution of the initial turbulent spectra and their effect on the ions. The results from the 2.5D simulations yield similar heating and acceleration rates for the ion species as the ones observed in 1.5D simulations \citep{Maneva:13b,Maneva:13a,Maneva:14}. The threshold value for the differential acceleration, $V_{\alpha p} = 0.5 V_\mathrm{A}$ agrees with previous results from 1.5D hybrid simulations with single monochromatic pump waves \citep{Maneva:14} and the power spectra of the magnetic field fluctuations shows that the initial parallel waves are evolving and growing predominantly parallel to the background magnetic field with less wave activity in the perpendicular direction. The combination of those effects suggests that the parametric instabilities and the turbulent evolution of low-frequency finite-amplitude Alfv\'en-cyclotron waves in 1.5D studies could serve as a starting point for understanding parallel wave propagation in two-dimensional turbulent problems. \\
Direct comparison between the nonlinear evolution of turbulence generated by parametrically unstable pump waves and a broad-band initial turbulent wave spectra based on a 1.5D hybrid simulation study shows similar spectral slopes for the late-stage magnetic field and density fluctuations. Nevertheless the resulting minor ion heating and differential acceleration are significantly different due to the different fluctuations at the early stage of their evolution, cf. with Fig. 3 from \citep{Maneva:13b}. This suggests that the two processes (of initial pump waves and initial turbulent wave spectra) should be considered complementary to each other, as they can be both realized in the solar wind, but have different consequences on the ions.\\
%We should note that 
For the plasma parameters considered here the linear Vlasov theory predicts much higher threshold value for the relative drift speed required for the onset of the two-streaming instability, $V_{\alpha p} > 1.8 V_\mathrm{A}.$ In this respect the observed deceleration of the $\alpha$ particles at $V_{\alpha p} > 0.5 V_\mathrm{A}$ is a nonlinear phenomenon, which requires higher-order analytical treatment or direct numerical simulations. In this study we have constructed a self-consistent initial wave spectra of parallel Alfv\'en-cyclotron waves and computed the related velocity fluctuations in the presence of differential streaming. We have estimated the apparent temperature anisotropies corresponding to the non-thermal velocity components generated by the initial wave spectra. We have demonstrated the effect of the differential streaming on the apparent temperature anisotropies for the minor ions. In a non-drifting plasma the $\alpha$ particles are strongly influenced by the initial spectra. They acquire high transverse bulk velocity fluctuations, with higher magnitude than the corresponding transverse component of the minor ion thermal speed. These fluctuations are significantly reduced in the presence of relative drifts. The protons are much less affected by the selected wave spectra and their non-thermal velocity component constitutes a fraction of the relevant proton thermal speed for all simulation cases. 
For the sake of clarity we should note that 2.5D simulations with higher initial relative drifts $V_{\alpha p} > 0.5 V_\mathrm{A}$ clearly show that the relative drift speed decreases in time. Furthermore, 1.5D simulations show similar decrease of the relative drift speed for $V_{\alpha p} \leq 1.8V_\mathrm{A}$  - i.e. throughout the entire linearly stable regime of the magnetosonic instability (without waves) for the given plasma parameters (see Fig. 9 from \citep{Maneva:14}). For larger drifts the system is prone to the linear theory streaming instability, which again causes the decrease of the relative drift speed in time.\\
Although one-dimensional simulations can provide useful hints for understanding the solar wind properties, they fail to capture other factors like oblique wave generation, anisotropic turbulence, plasma structures, etc. The solar wind expansion by construction is strongly influenced by the degrees of freedom in the system and its role can be stronger in two-dimensional systems. In addition, many of the observed solar wind magnetic fluctuations have oblique nature, whose scattering and interaction with the co-existing parallel waves require more than one spatial dimension. The present 2.5D simulation study provides a further step towards a proper treatment of the dynamic ion properties in the solar wind and the anisotropic evolution of the solar wind micro-turbulence, which cannot be captured by 1.5D modeling.\\
We find that the presence of relative drift speed prolongs the lifetime of the parallel waves at short wavelengths, as new waves with $|k_\parallel| \geq 0.35 \Omega_p/V_\mathrm{A}$ are generated and the wave power is replenished after the initial spectra is damped. %which are depleted in the non-drifting case due to resonant absorption by the minor ions. 
This implicates that the turbulent evolution of the plasma waves and the nonlinear wave-wave interactions are affected by the differential streaming. In addition the relative drifts enter the cyclotron damping resonant condition and Doppler-shift the frequency of the Alfv\'en-cyclotron waves, required for resonant wave-particle interactions with the minor ions. The selected initial spectra is far from resonance with the protons for all values of the relative drift speeds considered in this study. In addition the direct turbulent cascade is not strong enough to bring sufficient wave power at the proton-cyclotron scales during the short (kinetic) timescale of the simulations. Hence there is no perpendicular heating for the protons at any time throughout the simulations. The differential streaming also suppresses the parallel wave generation at large wavelengths $|k_\parallel| < 0.125 \Omega_p/V_\mathrm{A}$ and facilitates the oblique mode generation with enhanced energy transfer in the perpendicular direction. In both drifting and non-drifting plasmas the oblique waves appear confined within the range of $|k_\perp| \in [0,0.3] \Omega_p/V_\mathrm{A},$ with most of their power concentrated in $|k_\perp| \in [0-0.1]\Omega_p/V_\mathrm{A}.$ \\
The close-to-realistic values of the solar wind expansion factor in the inner heliosphere used here, $\varepsilon t_0 = 10^{-4}$, lead to only a few percent change in the evolution of the differential streaming on the temporal scales considered here. Still, the expansion plays significant role in decreasing the ion temperature anisotropies and the perpendicular ion temperatures -- up to 20\% for the $\alpha$ particles. As the minor ions are preferentially heated by the existing waves in the system, their perpendicular temperature remains higher than the double adiabatic prediction. The protons on the other hand experience no wave heating and their cooling in time is practically given by the CGL model. The slight additional cooling comes from the lack of exact energy conservation in the system, where the total electromagnetic energy is preserved within $14\%$, the kinetic energy of the protons is preserved with $1.5\%$ and the kinetic energy for the alpha particles is preserved within $11\%$. Although the energy is quasi-conserved, there is still some energy loss in the system as all the above energies slightly decrease in time. If we account for this numerical effect we would expect even higher heating for the minor ions.\\
We should note that the gradual solar wind expansion considered here would have stronger net effect as we follow the plasma evolution over a longer time interval. Assuming expansion factor of $\varepsilon t_0 = 5\times10^{-4}$ already leads to a substantial reduction of the relative drift speed (up to $40\%$) and the ion temperature anisotropies (up to 51\%), as visible from Table~1. The faster expansion also changes the nonlinear evolution of the initial wave spectra, affecting the wave-particle interactions and the resulting shape of the ion velocity distribution functions. Since the wave-particle interactions operate much faster than the expansion time-scales, though, a pure increase in the expansion factor might not properly describe the plasma evolution. A self-consistent description of the evolution of the turbulent fast solar wind plasma requires long simulation times and is computationally very expensive even in 2.5D hybrid kinetic models. It is out of the scope of this paper to investigate the turbulent cascade at very large MHD scales. However the present study is able to capture the dynamics at the ion scales and demonstrate the back-reaction of the ions on the cascade processes and the evolution of the turbulent wave spectra at the intermediate ion scales and some limited part of the larger MHD spectra. \\
\begin{figure}
\centering
\noindent\includegraphics[width=0.46\textwidth]{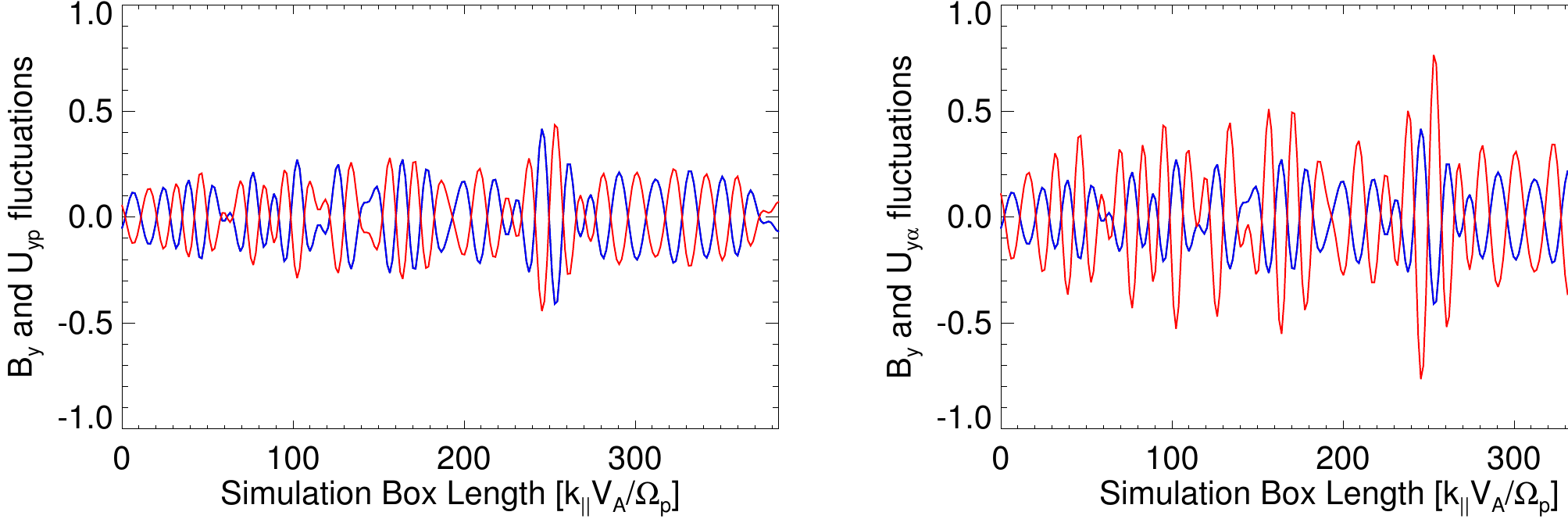}
\noindent\includegraphics[width=0.46\textwidth]{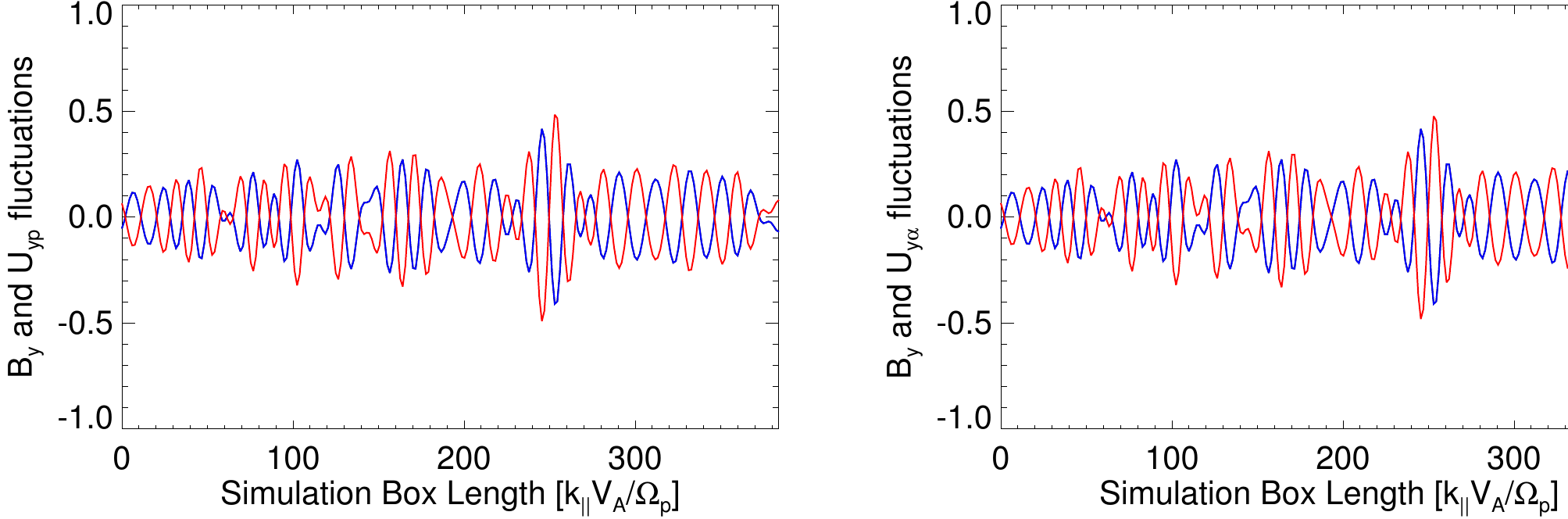}
\noindent\includegraphics[width=0.46\textwidth]{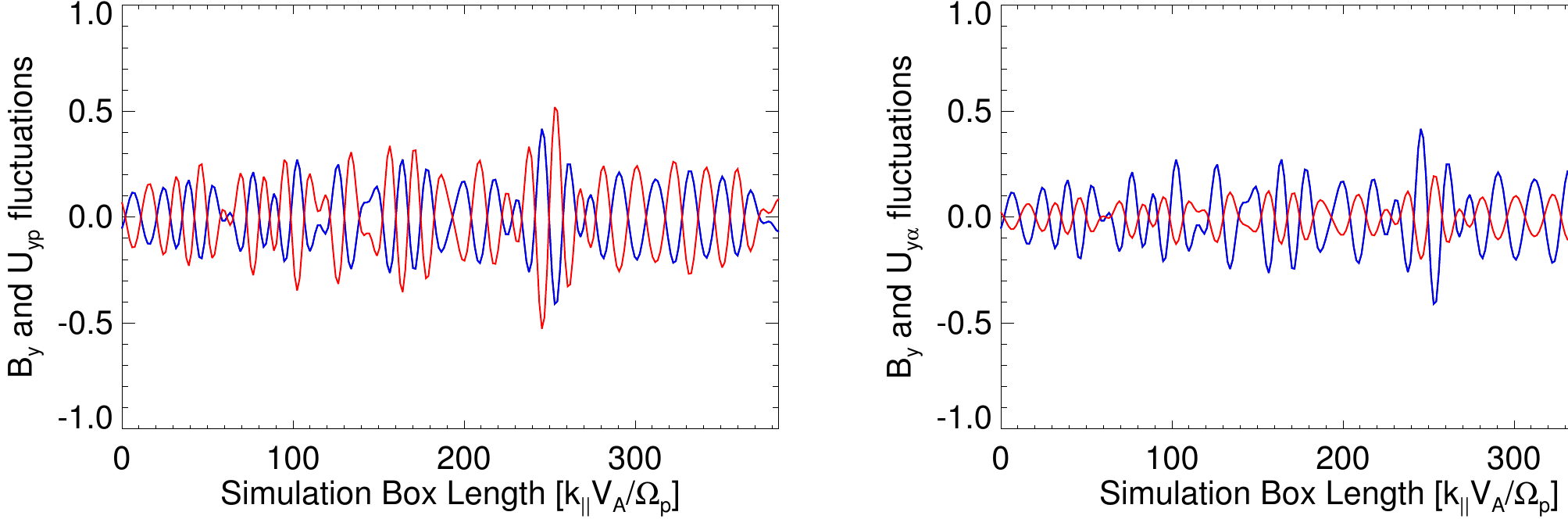}
%\noindent\includegraphics[width=0.48\textwidth]{spect_epa_realsp_16modes_alf1_b02_run39_by_uy_vn_v2.pdf}
%\noindent\includegraphics[width=0.48\textwidth]{spect_epa_realsp_16modes_alf1_b02_run27_by_uy_vn_v2.pdf}
%\noindent\includegraphics[width=0.48\textwidth]{spect_epa_realsp_16modes_alf1_b02_run28_by_uy_vn_v2.pdf}
\caption{Magnetic field (blue lines) and transverse velocity fluctuations (red lines) for the protons and $\alpha$ particles, corresponding to the initial broad-band spectra of Alfv\'en-cyclotron waves. The plot shows the strong dependence of the induced transverse velocity fluctuations for the minor ions on the value of the initial relative drift speed.}
\label{fig:aniso_dr00}
\end{figure}
\begin{figure}
\centering
%\noindent\includegraphics{aniso_2runs_run14_run15_col.pdf}
%$\varepsilon=0.5*10^{-4}$
\noindent\includegraphics[width=0.48\textwidth]{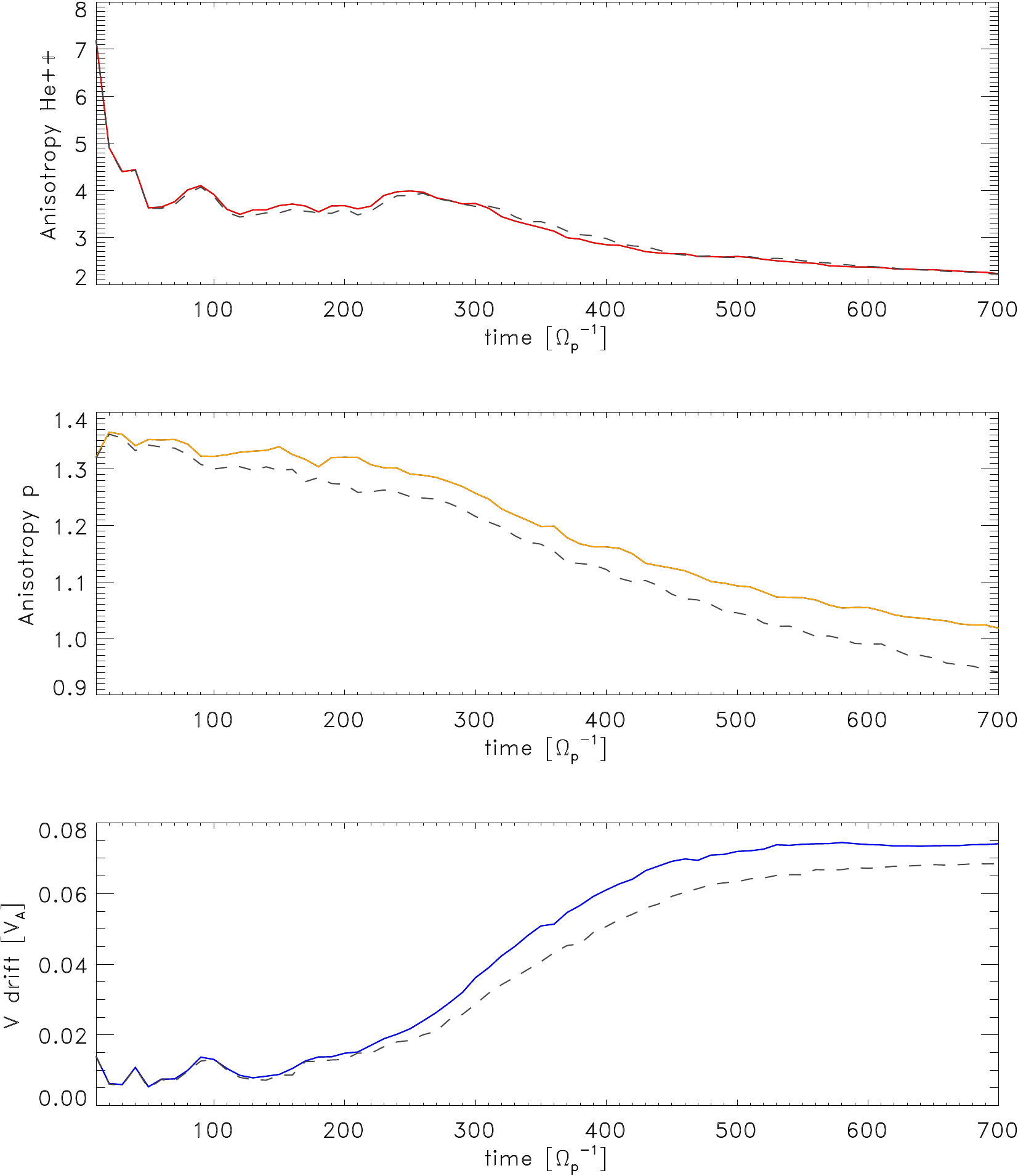}
\caption{Temporal evolution of the ion temperature anisotropies and relative drift speed, showing the generation of differential streaming between protons and $\alpha$ particles. The solid lines indicate simulations without expansion and the dashed lines show the results when a gradual expansion with $\varepsilon t_0=10^{-4}$ was considered. The apparent initial anisotropies are caused by the waves.}
\label{fig:aniso_dr0}
\end{figure}
\begin{figure}
\centering
\noindent\includegraphics[width=0.48\textwidth]{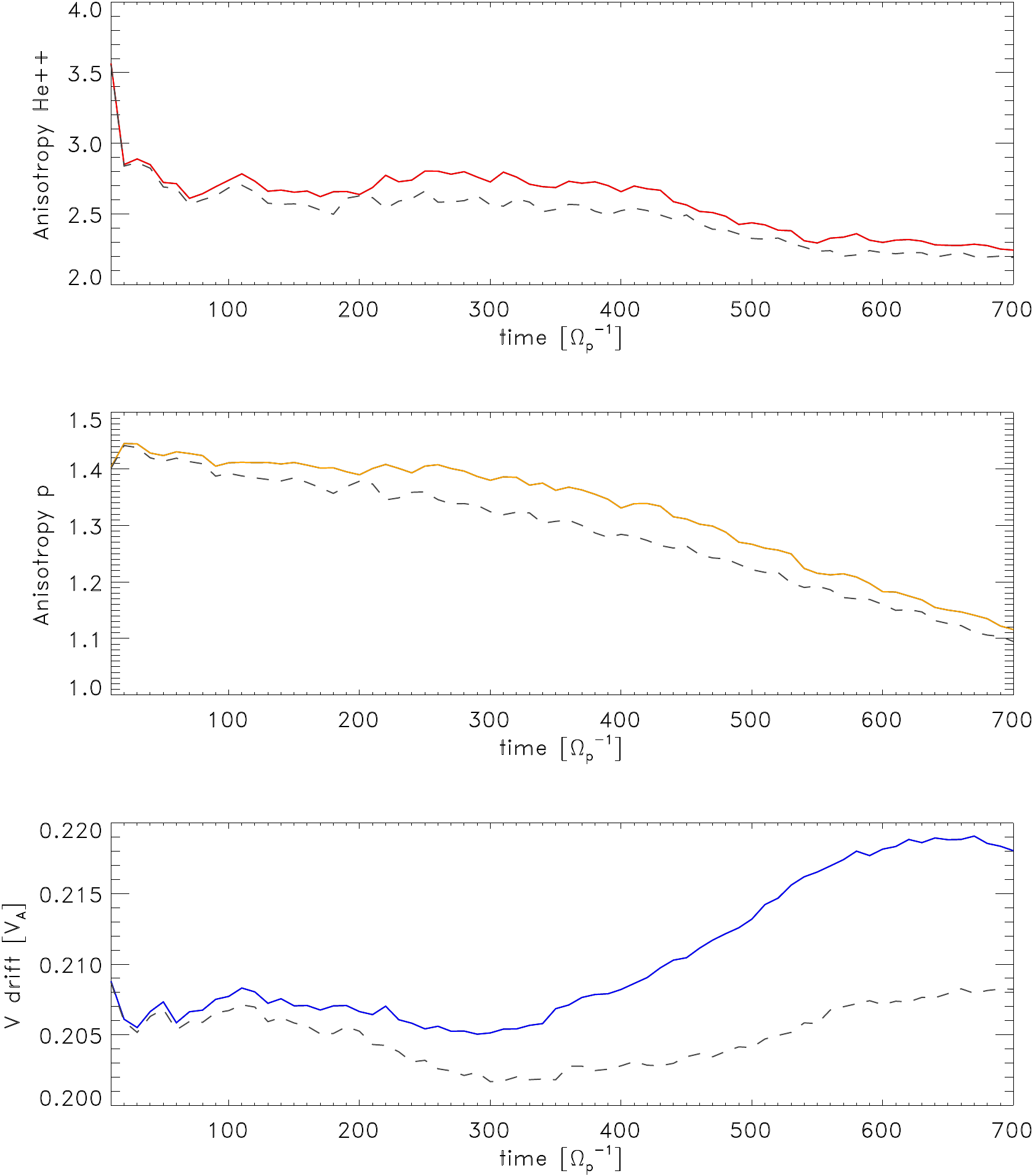}
\caption{Temporal evolution of the ion temperature anisotropies and relative drift speed for the case of differentially streaming $\alpha$ particles with $V_{\alpha p} = 0.2 V_\mathrm{A}$. The dashed lines denote the ion bulk properties in the case of a gradual expansion with $\varepsilon t_0=10^{-4}$.}
\label{fig:aniso_dr02}
\end{figure}

\begin{figure}
\centering
\includegraphics[width=0.48\textwidth]{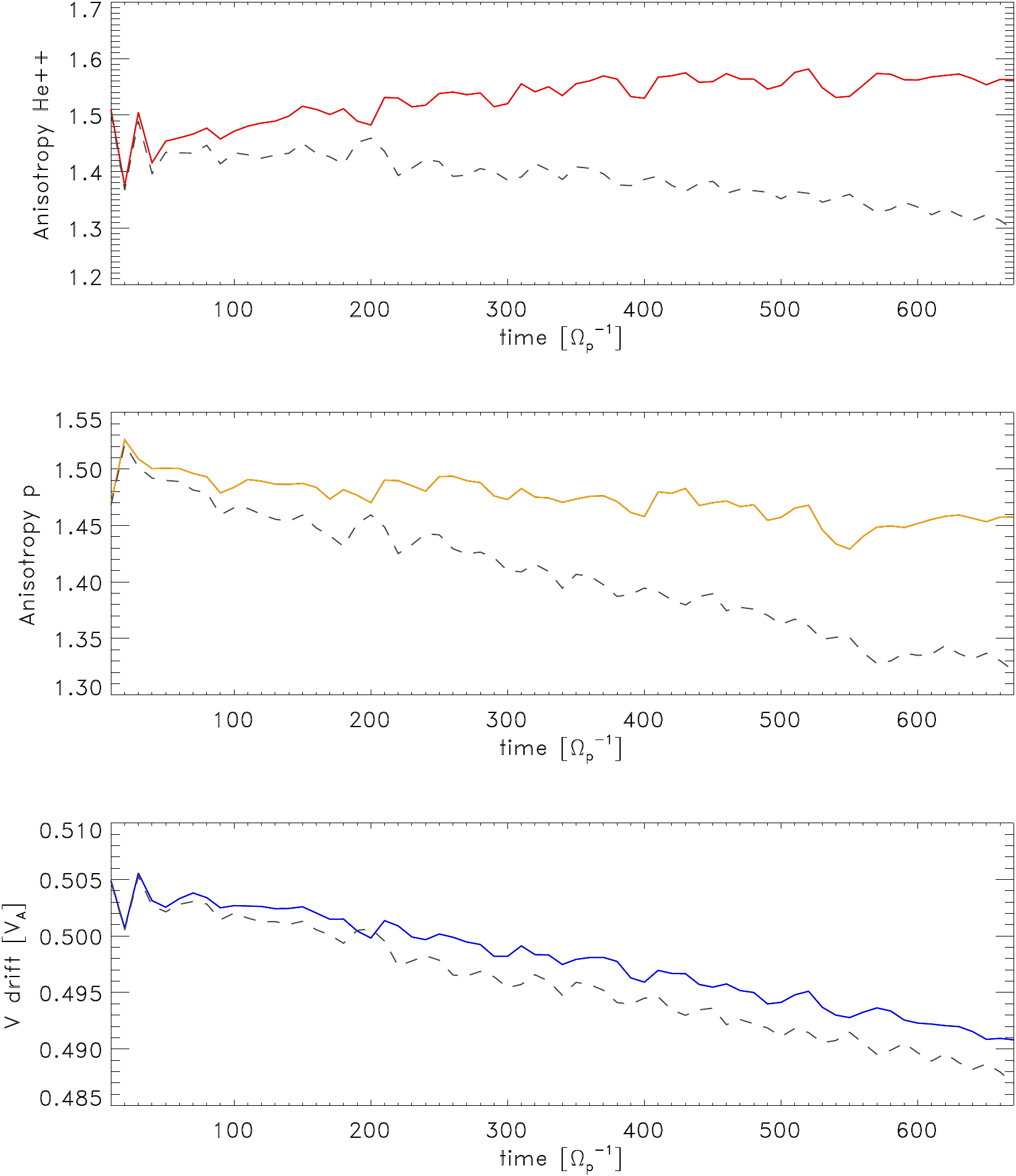}
\caption{Temporal evolution of the ion temperature anisotropies and relative drift speed for the case of differentially streaming $\alpha$ particles with $V_{\alpha p} = 0.5 V_\mathrm{A}$. The solid lines indicate simulations without expansion and the dashed lines denote the results with $\varepsilon t_0=10^{-4}$.}
\label{fig:aniso_dr05}
\end{figure}

\begin{figure}
\centering
\includegraphics[width=0.47\textwidth]{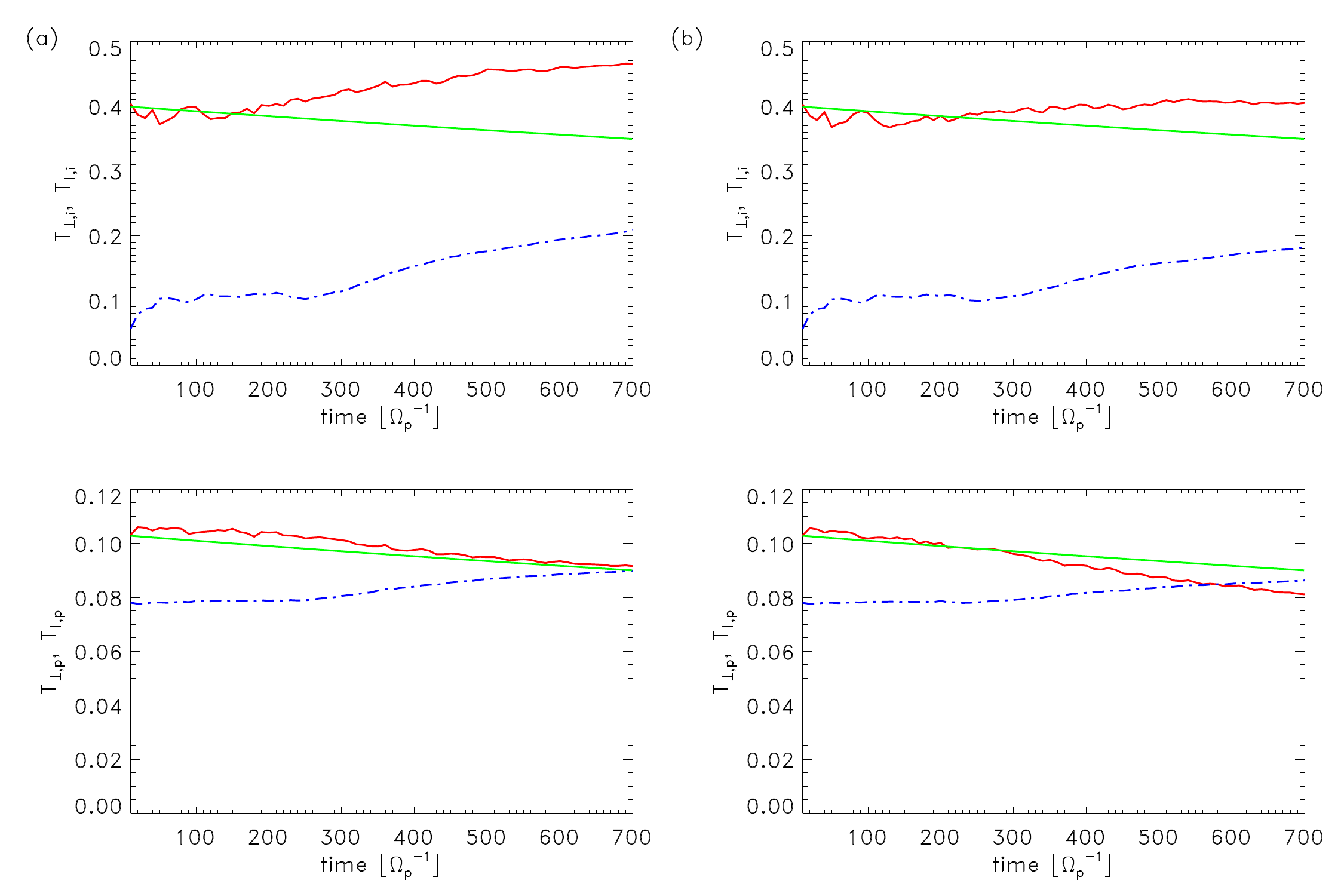}
\caption{Temporal evolution of the parallel and perpendicular components of the ion temperatures for initially non-drifting distributions. Panels (a) shows a non-expanding case. Panel (b) illustrates the expansion with $\varepsilon t_0=10^{-4}$. The top rows describe the $\alpha$ particles, whereas the bottom ones depict the protons. The solid red lines denote the perpendicular and the dash-dotted blue lines denote the parallel temperature components. The solid green lines denote the perpendicular temperature decrease as predicted by the CGL model.}
\label{fig:temp_dr0}
\end{figure}

\begin{figure}
\centering
\includegraphics[width=0.47\textwidth]{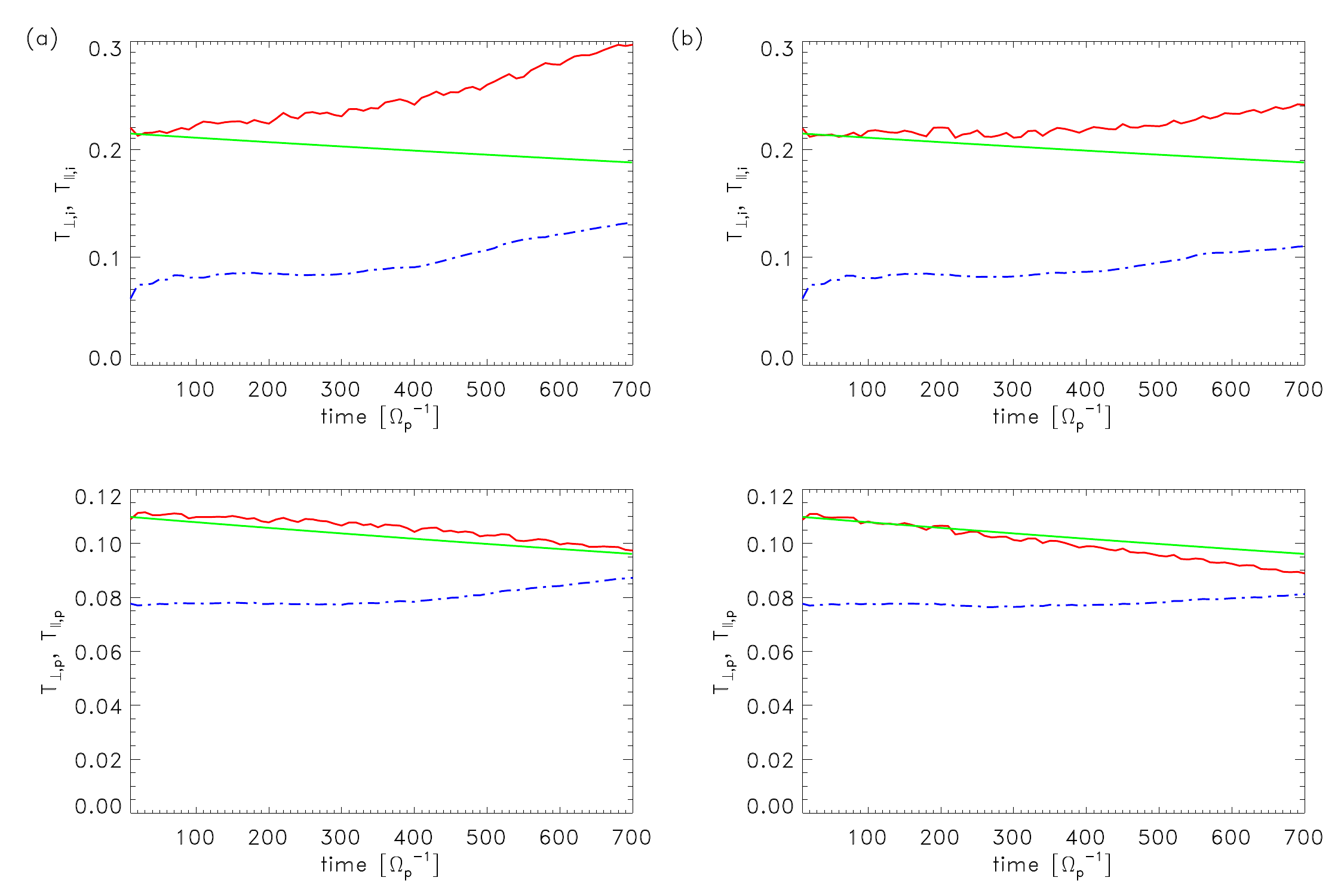}
\caption{Temporal evolution of the individual ion temperatures for the case of initially drifting plasma with $V_{\alpha p} = 0.2 V_\mathrm{A}.$ Similar to Figure~\ref{fig:temp_dr0}, the right panel shows the effect of a gradual expansion with $\varepsilon t_0=10^{-4}$.}
\label{fig:temp_dr02}
\end{figure}

\begin{figure}
\centering
\includegraphics[width=0.47\textwidth]{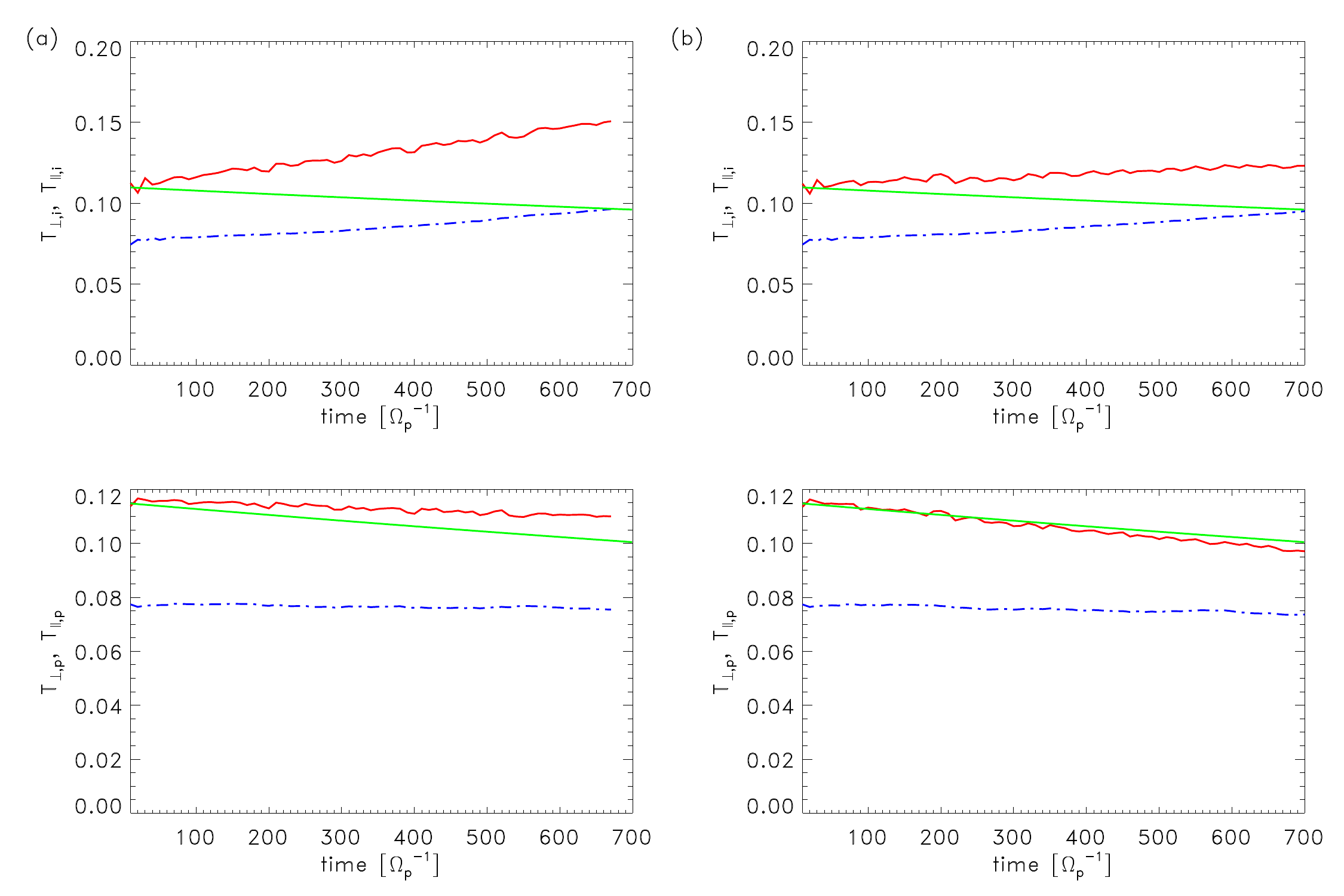}
\caption{Temporal evolution of the individual ion temperatures for the case of initially drifting plama with $V_{\alpha p} = 0.5 V_\mathrm{A}.$ The right panel shows the effect of the gradual expansion with $\varepsilon t_0=10^{-4}$. The notations are the same as on Figure~\ref{fig:temp_dr0} and Figure~\ref{fig:temp_dr02}.}
\label{fig:temp_dr05}
\end{figure}

\begin{figure}
\centering
\includegraphics[width=0.24\textwidth]{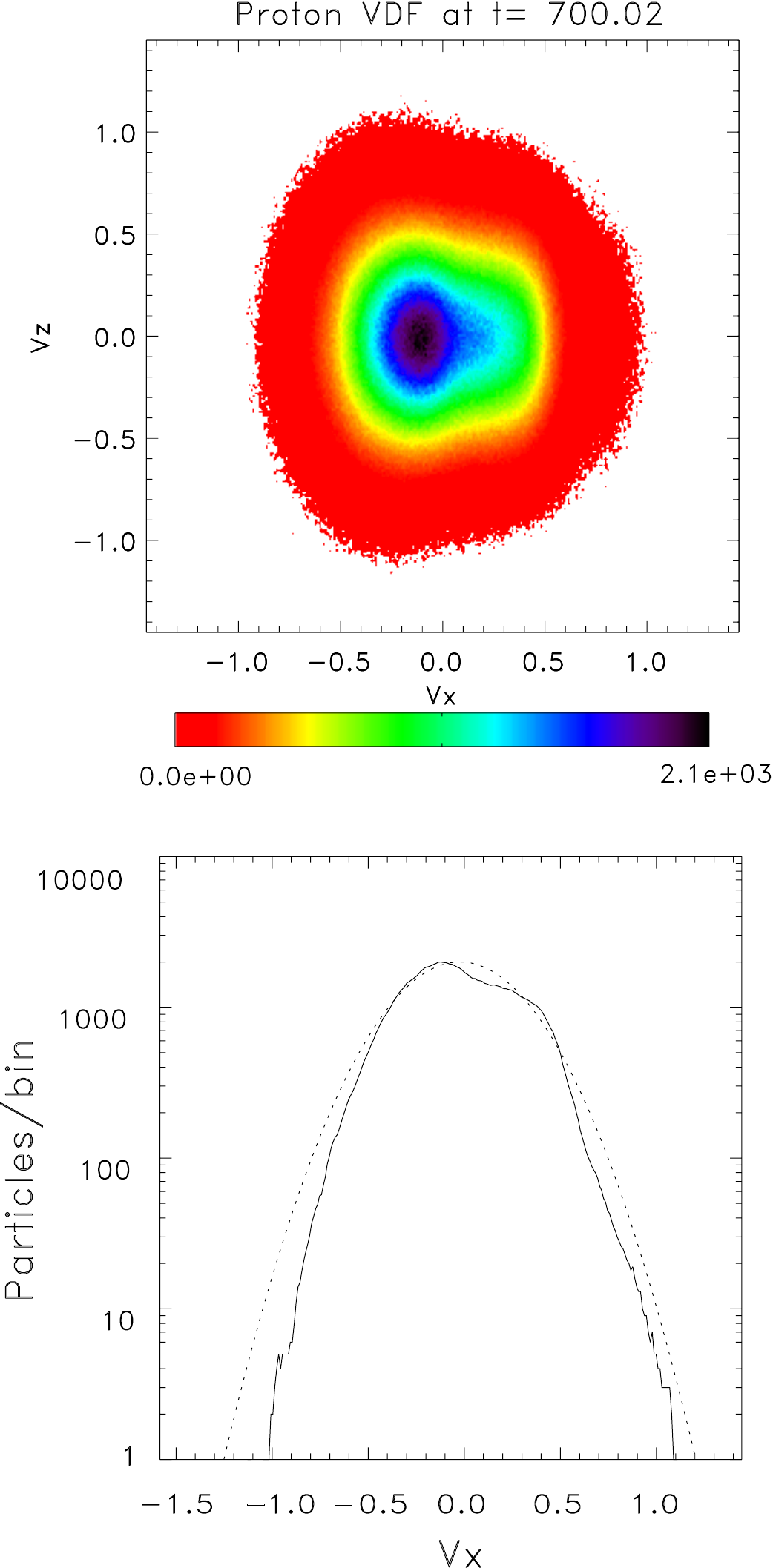}
\includegraphics[width=0.24\textwidth]{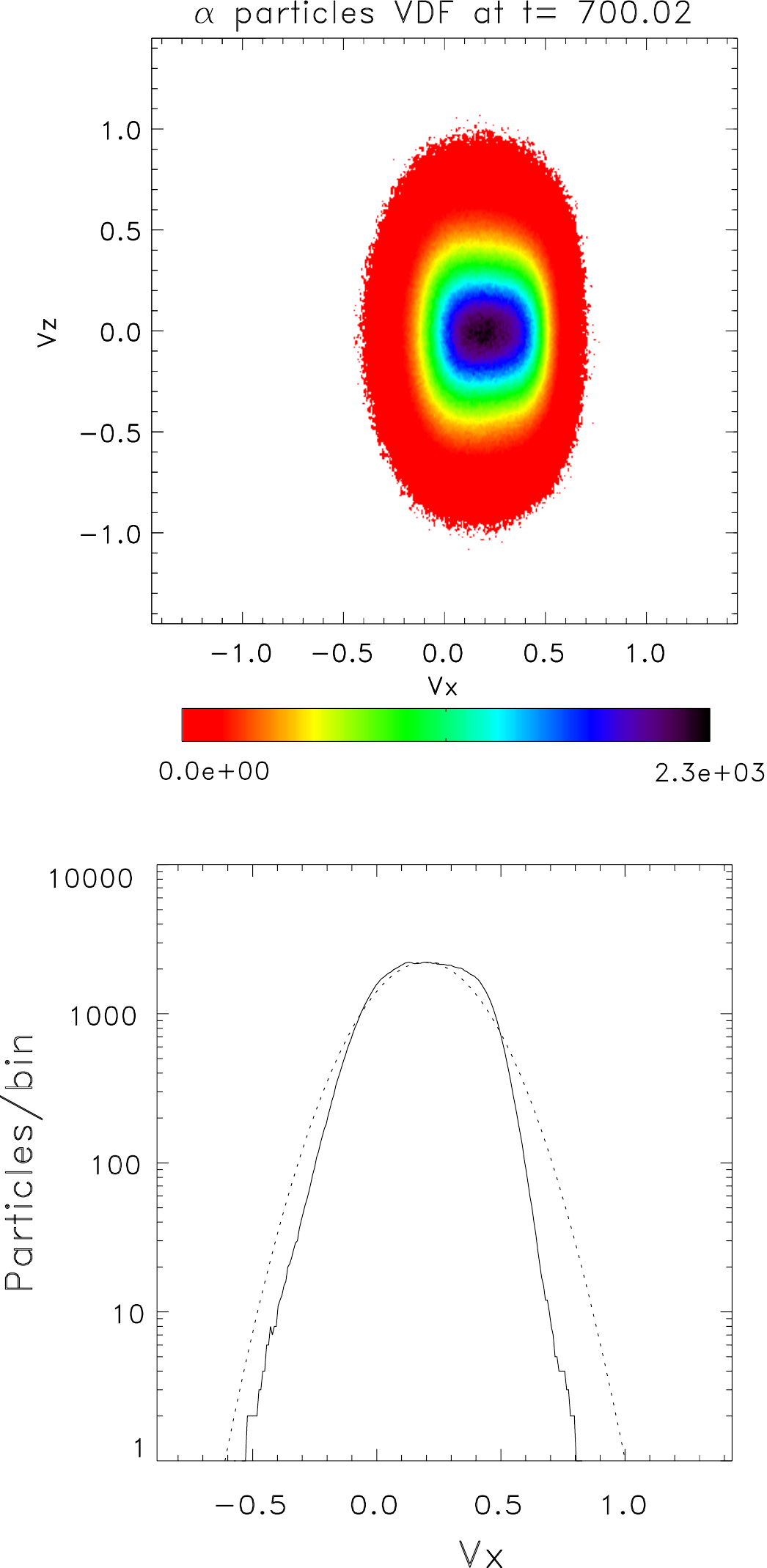}
%$\varepsilon=10^{-4}$
\caption{Snapshot of the two-dimensional velocity distribution functions for protons (left) and $\alpha$ particles (right) for the case of initially drifting isotropic and isothermal plasma with $V_{\alpha p} = 0.2 V_\mathrm A.$ The reduced distributions in the lower panel show the number of particles as a function of the parallel component of the ion thermal speeds. A best-fit Maxwellian velocity distribution is overplotted with the dotted line.}
\label{fig:VDF_dr02}
\end{figure}

\begin{figure}
\centering
 \includegraphics[width=0.3\textwidth, height=0.3\textheight, angle=90]{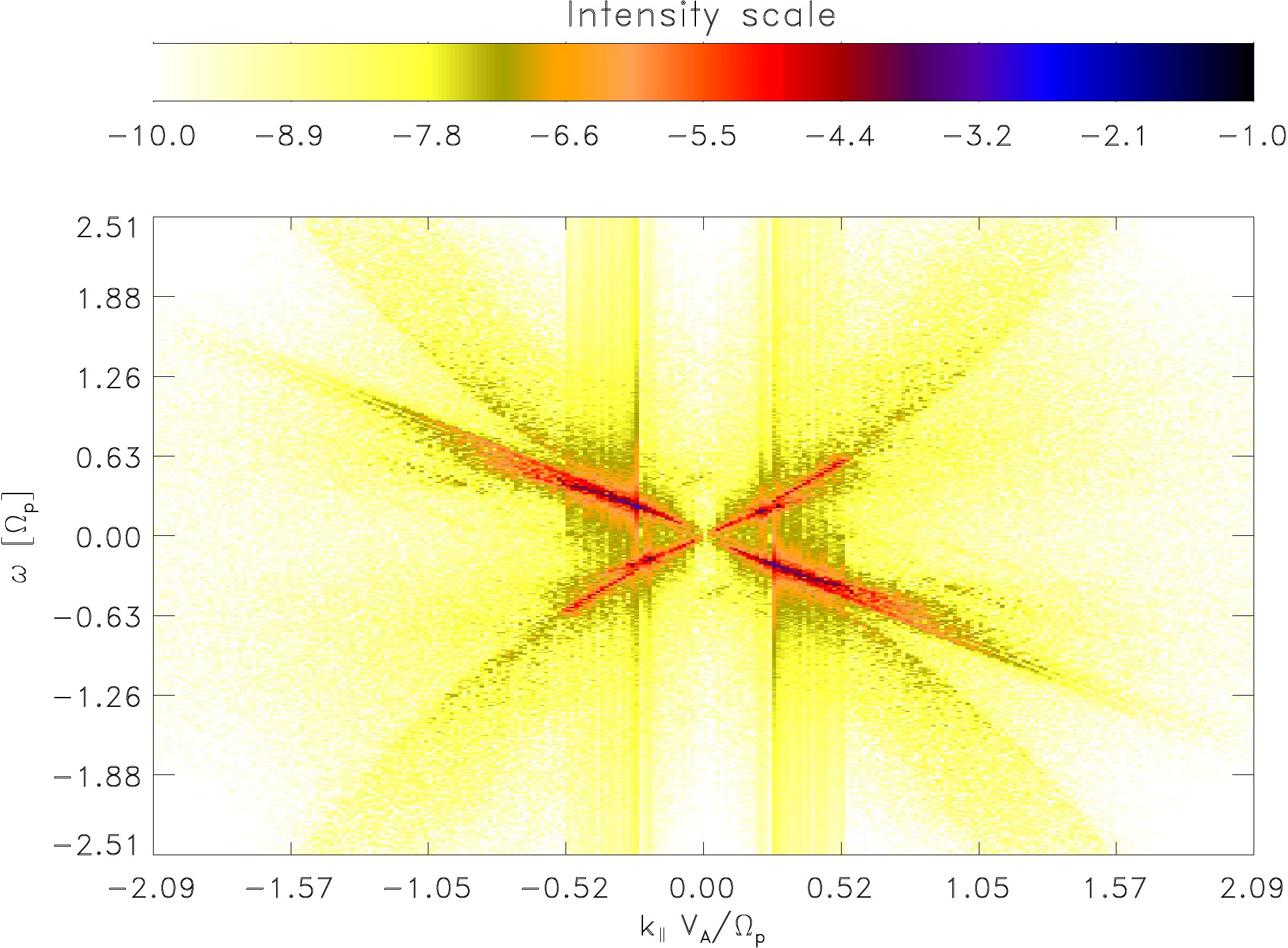}
 \hspace{2pt}
 \includegraphics[width=0.3\textwidth, height=0.3\textheight, angle=90]{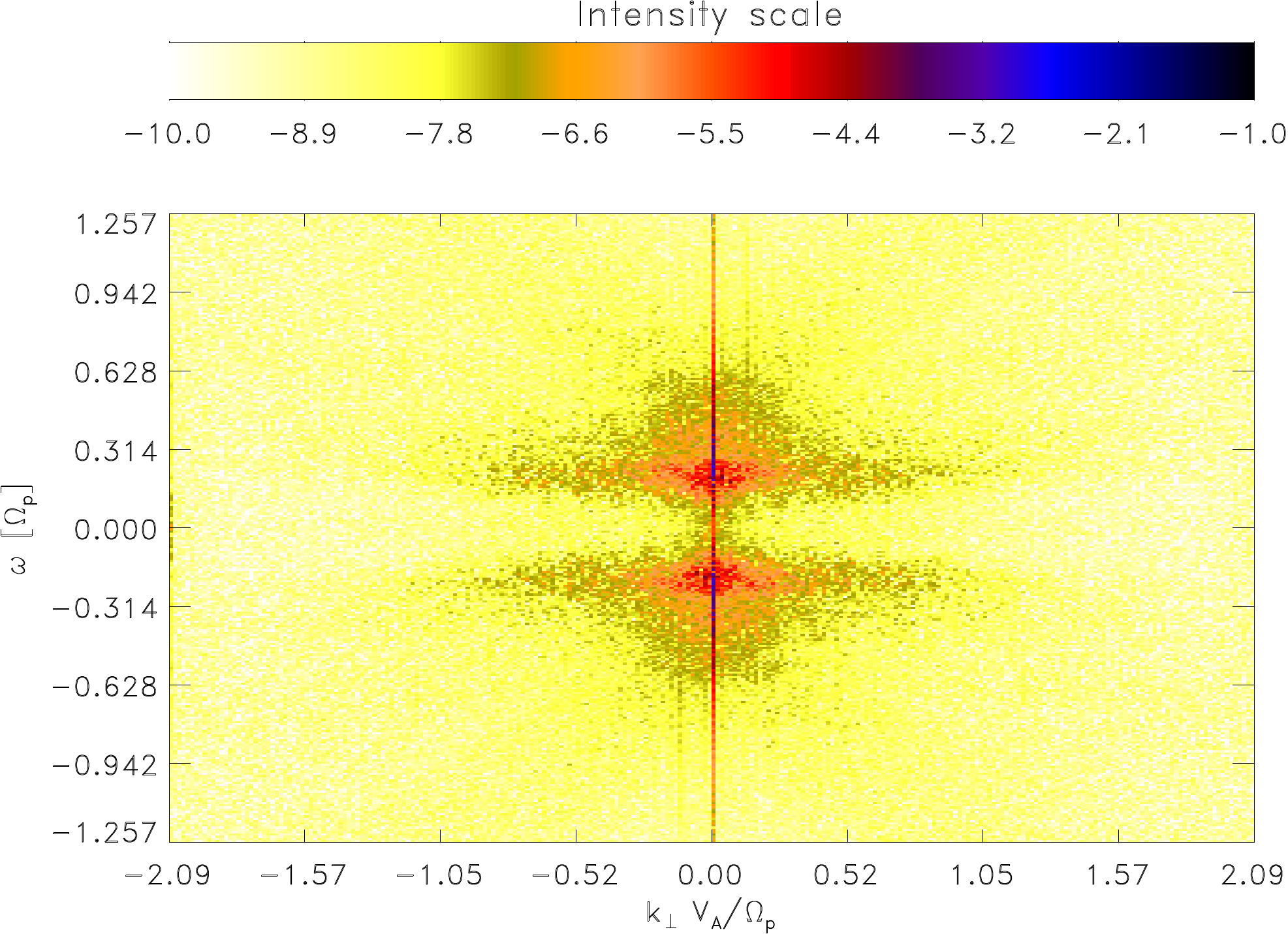}
\caption{Dispersion diagram $\omega$ vs. $k$, representing the wave intensity as a function of wave frequency vs parallel (up) and perpendicular (bottom) wave-number space. The graph corresponds to the case of non-drifting initial distributions.}
\label{fig:spectra_run14_par_per}
\end{figure}

\begin{figure}
\centering
\includegraphics[width=0.3\textwidth, angle=90]{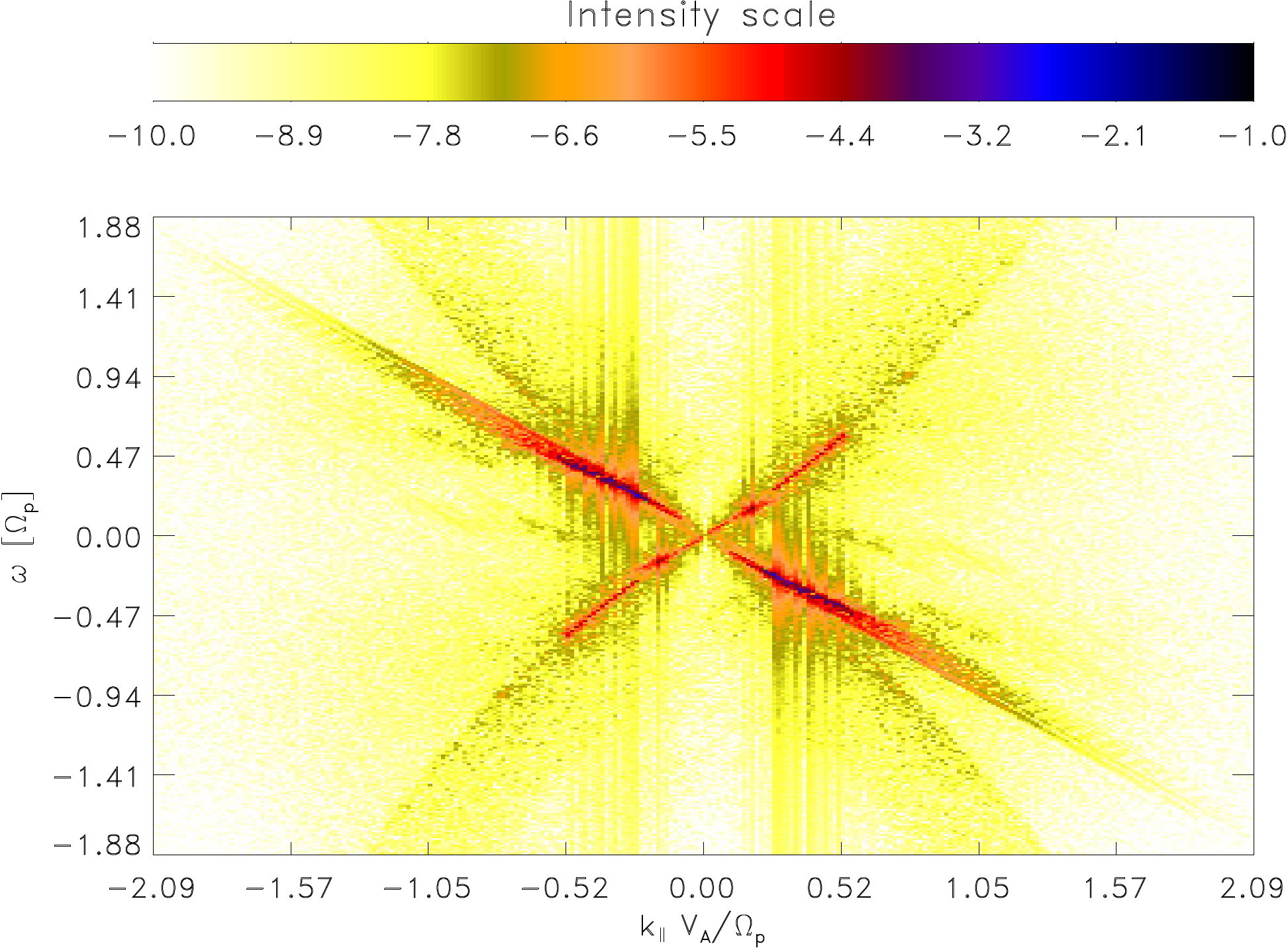}
\includegraphics[width=0.3\textwidth, angle=90]{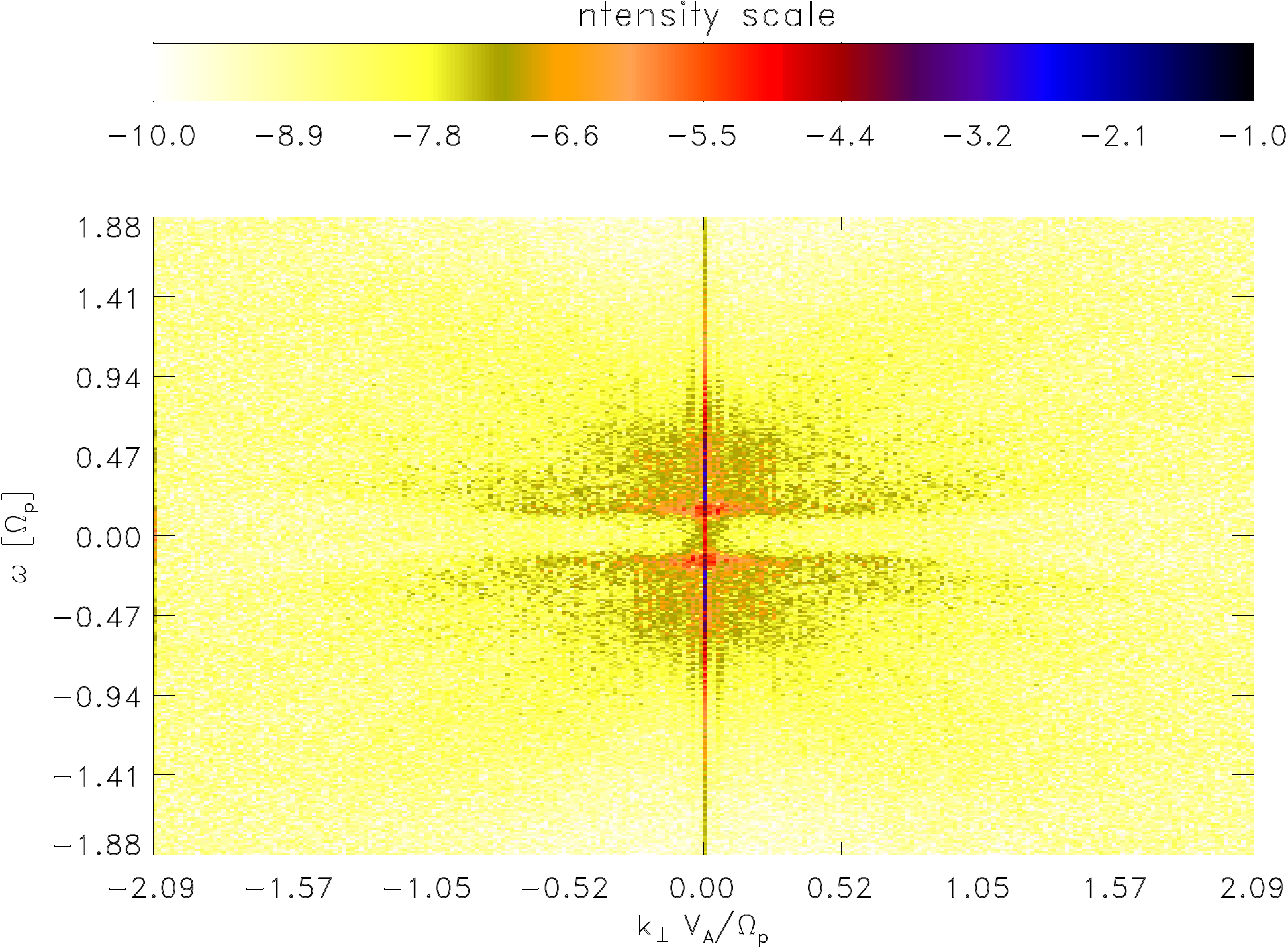}
\caption{Wave intensity as a function of wave frequency vs parallel (up) and perpendicular (bottom) wave-number for the case of a drifting non-expanding plasma with $V_{\alpha p} =0.2 V_\mathrm{A}$.}
\label{fig:spectra_run15_par_per}
\end{figure}

\begin{figure}
\centering
\includegraphics[width=0.47\textwidth]{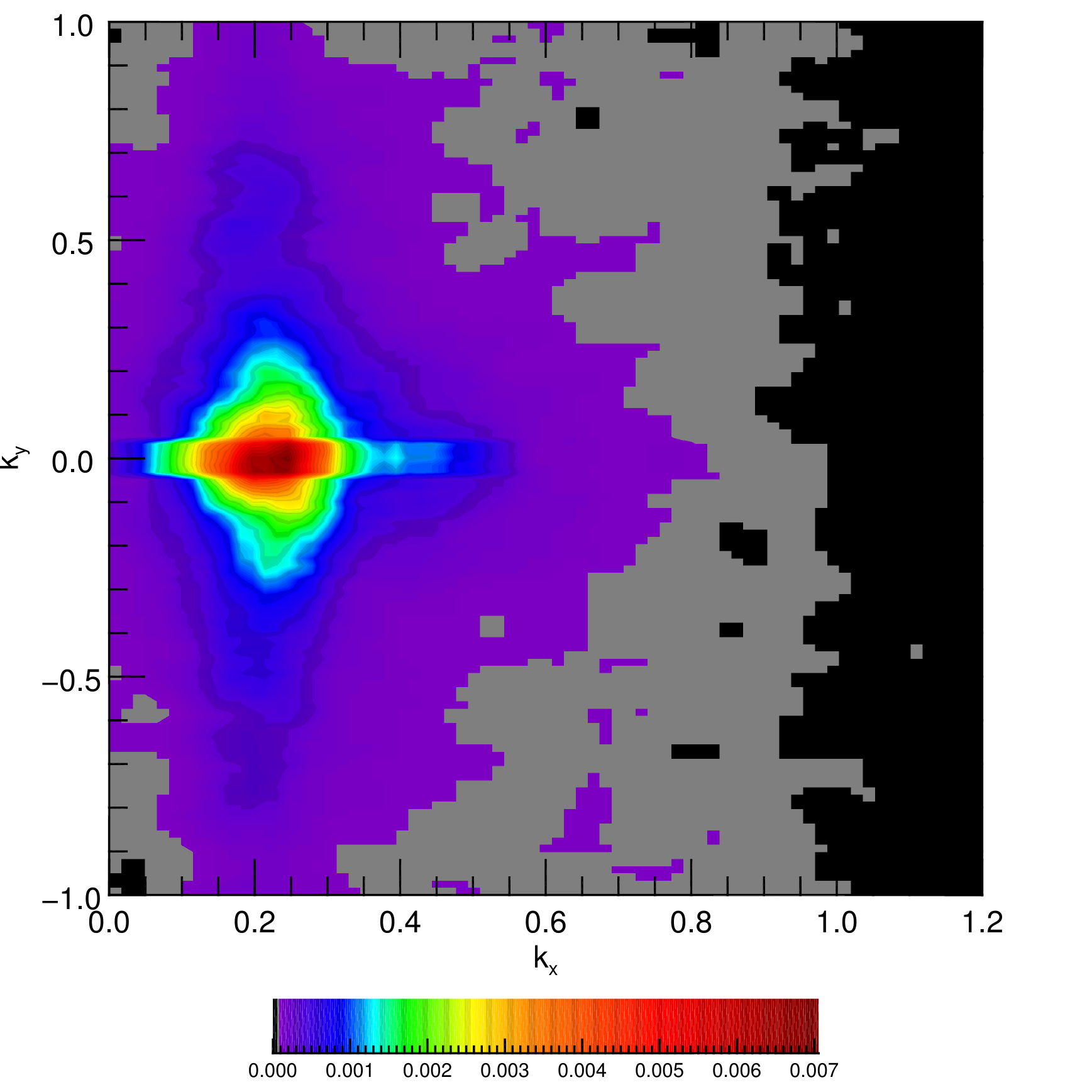}
\caption{Wave power as a function of the parallel vs. perpendicular wave-number for the case of a non-drifting non-expanding plasma at the end stage of the simulations $\Omega_pt=700$.}
\label{fig:kpar_kper_dr0}
\end{figure}

\begin{figure}
\centering
\includegraphics[width=0.47\textwidth]{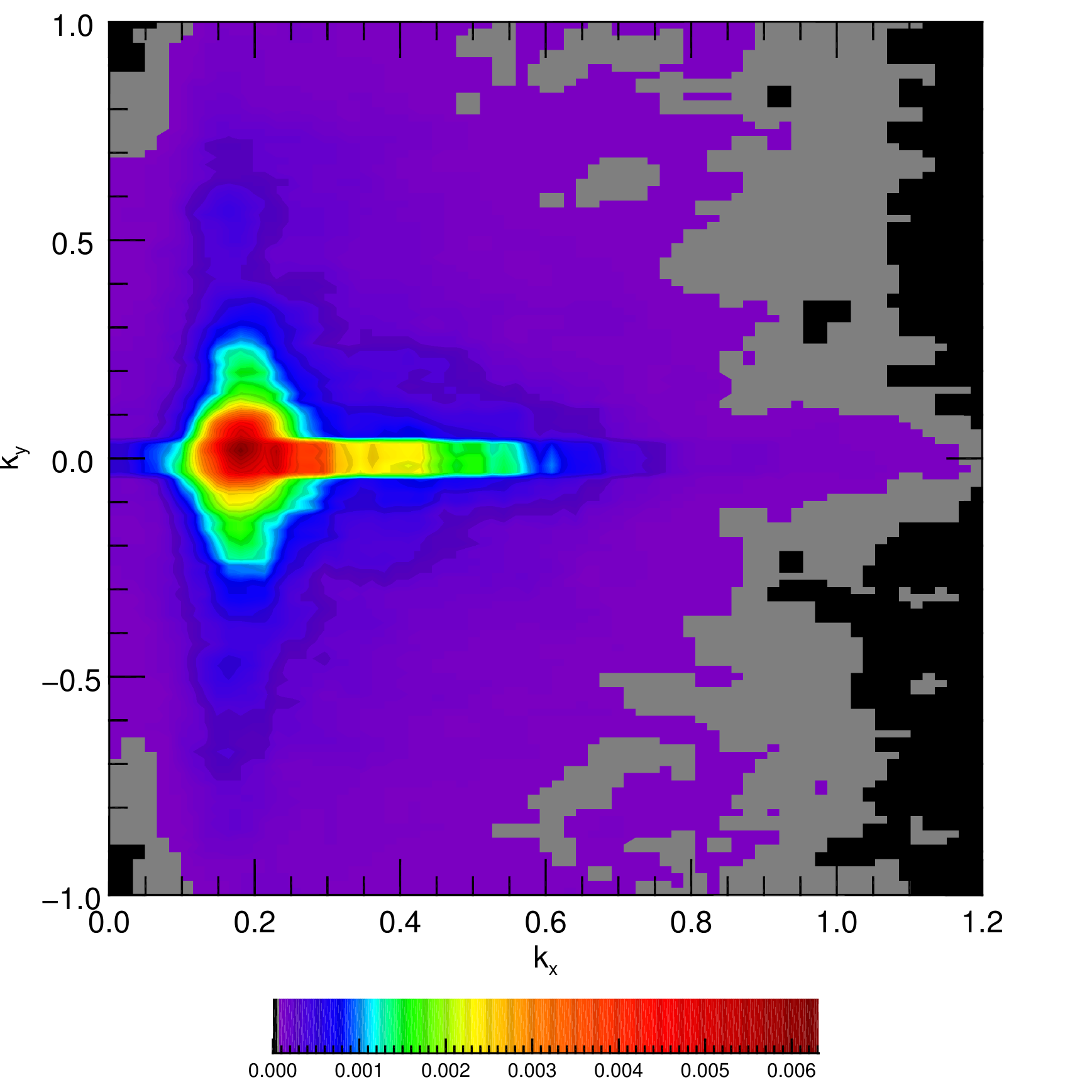}
\caption{Wave power as a function of the parallel vs. perpendicular wave-number at $\Omega_pt=700$ for the case of a drifting non-expanding plasma with $V_{\alpha p} =0.2 V_\mathrm{A}$.}
\label{fig:kpar_kper_dr02}
\end{figure}

\begin{table*}
\tiny
\label{runparams}
\caption{Summary of the parameters used in the 2.5D hybrid simulation study. In column 2 and 3 we present the range of the normalized wave frequencies and wave-numbers, followed by the expansion parameter $\varepsilon t_0$ in column 4. Next are the ratios of the average of the wave-induced ion perpendicular bulk velocity fluctuations to the perpendicular component of the ion thermal speeds for each species, given in columns 5 and 6. The last four columns show the initial $V_{\alpha p, \mathrm{in}}$ and the final stage values $V_{\alpha p, \mathrm{fin}}$ of the relative drift speed, followed by the ion temperature anisotropies for protons and $\alpha$ particles, calculated at the end of the simulations.}
%\begin{center}
\centering
  \begin{tabular}{|c|c|c|c|c|c|c|c|c|c|c|}
    \hline
    Case~\# & $\omega_0$ [$\Omega_p$] & $k_0$ [$\Omega_p/V_\mathrm{A}$] & Exp. param. $\varepsilon t_0$ & $<V_{\perp \alpha}> /v_{\mathrm{th}, \perp \alpha}$ & $<V_{\perp p}>/v_{\mathrm{th}, \perp p}$ & $V_{\alpha p, \mathrm{in}} [V_{\mathrm A}]$ & $V_{\alpha p, \mathrm{fin}}[V_{\mathrm A}]$ & $T_{\perp p}/T_{\parallel p}$ & $T_{\perp \alpha}/T_{\parallel \alpha}$ \\ \hline
    1 &  [0.21-0.36] &  [0.26-0.52]  & 0 &1.84 &0.53 & 0 & 0.074 & 1.2& 2.2\\ %run1 (for run14 - 0.066)
    2 &  [0.21-0.36] & [0.26-0.52] & $5\times10^{-4}$ &1.84 &0.53 & 0 & 0.068 & 0.94 & 2.2\\ %run2
    3 &  [0.21-0.36] & [0.26-0.52] & $10^{-4}$ &1.84 &0.53 & 0 & 0.05 &0.7 &2.1 \\ %run 2_old, run15 - 0.056
    4 &  [0.22-0.386] & [0.26-0.52]  & 0 &1.15 & 0.58 & 0.2 & 0.218 & 1.14 & 2.25\\ %run3, run27
    5 &  [0.22-0.386] & [0.26-0.52]  & $5\times10^{-4}$ &1.15 &0.58 &0.2 & 0.208 & 1.09 &2.2 \\ %run4, run30
    6 &  [0.22-0.386] & [0.26-0.52]  & $10^{-4}$ &1.15 &0.58 &0.2 & 0.174 & 0.84& 1.3\\ %run30
    7 &  [0.22-0.39] & [0.26-0.52] & 0 & 0.42 & 0.60 & 0.5 & 0.491 & 1.46& 1.56\\%run28
    8 &  [0.22-0.39] & [0.26-0.52]  & $5\times10^{-4}$ &0.42 &0.60 & 0.5 & 0.488 &1.33 &1.31\\ %run29
    9 &  [0.22-0.39] & [0.26-0.52]  & $10^{-4}$ &0.42 &0.60 & 0.5 & 0.477 & 0.96& 0.78\\ %run29
%2 &  [0.21-0.36] & [0.26-0.52] & $10^{-4}$ &1.84 &0.53 & 0 & 0.056 & &\\ %run 2 new
%4 &  [0.22-0.386] & [0.26-0.52]  & $10^{-4}$ &1.15 &0.58 &0.2 & 0.174& &\\ %run 4 new
%6 &  [0.22-0.39] & [0.26-0.52]  & $10^{-4}$ &0.42 &0.60 & 0.5 & 0.477& &\\ %run 6
%7 &  [0.21-0.36] & [0.26-0.52]  & 0 & 0 & ...\\ %run16
% 8 & [0.21-0.36] & [0.26-0.52] & $ 5*10^{-4} & 0 & ...\\%run17
   \hline
  \end{tabular}
%\tablenotetext{a}{Footnote text here.}
%\end{center}
\end{table*}

\begin{acknowledgements}
This work was supported by NASA, grant NNX10AC56G. Y.~G.~Maneva would like to thank the $F^{+}$ fellowship at KU Leuven for the partial support. A.~F.~Vi{\~n}as would like to thank the Wind/SWE MO/DA grant for the support. Fruitful discussions with P. Hunnana are highly appreciated.
\end{acknowledgements}

%\bibliographystyle{aa}
%\bibliography{Maneva_2dhyb}

\end{document}